\documentclass[aps,prd,twocolumn,superscriptaddress,longbibliography,nofootinbib]{revtex4-2}
\usepackage[utf8]{inputenc}
\usepackage{color}
\usepackage{graphicx}

\usepackage{amsmath,amssymb,amsfonts}

\def\prl{Phys. Rev. Lett.}
\def\prd{Phys. Rev. D}

\newcommand\longcomment[1]{}

\begin{document}

\title{Critical phenomena in the gravitational collapse of electromagnetic dipole and quadrupole waves}

\author{Maria F.~Perez Mendoza}

\affiliation{Department of Physics and Astronomy, Bowdoin College, Brunswick, ME 04011}

\author{Thomas W.~Baumgarte}

\affiliation{Department of Physics and Astronomy, Bowdoin College, Brunswick, ME 04011}

\begin{abstract}
We report on critical phenomena in the gravitational collapse of electromagnetic waves.  Generalizing earlier results that focused on dipole electromagnetic waves, we here compare with quadrupole waves in axisymmetry.  We perform numerical simulations of dipole and quadrupole wave initial data, fine-tuning both sets of data to the onset of black hole formation in order to study the critical solution and related critical phenomena.  We observe that different multipole moments have different symmetries, indicating that the critical solution for electromagnetic waves cannot be unique, at least not globally.  This is confirmed in our numerical simulations: while dipole data lead to a single center of collapse, at the center of symmetry, quadrupole data feature two separate centers of collapse on the symmetry axis, above and below the center of symmetry -- reminiscent of similar findings reported for critical collapse of vacuum gravitational waves.  While the critical solution for neither the dipole nor the quadrupole data is exactly self-similar, we find that their approximate echoing periods appear to differ, as do the critical exponents.  We discuss whether the centers of collapse found for dipole and quadrupole data might all have the same properties, which would suggest a ``local uniqueness" of the critical solution.  Instead, we provide some evidence -- including the differing echoing periods and critical exponents -- suggesting that the critical solutions are distinct even locally.  We speculate on the implications of our findings for critical phenomena in the collapse of vacuum gravitational waves, which share with electromagnetic waves the absence of a spherically symmetric critical solution.
\end{abstract}

\maketitle

\section{Introduction}
\label{sec:intro}

Critical phenomena in gravitational collapse were first reported in the seminal work of Choptuik \cite{Cho93}, who studied massless scalar fields, minimally coupled to gravity, in spherical symmetry.   Considering families of initial data parametrized by some parameter $\eta$, Choptuik distinguished {\em subcritical} data, which ultimately disperse to infinity, leaving behind flat space, from {\em supercritical} data, which collapse to form a black hole.  Reminiscent of similar effects in other fields of physics and beyond, Choptuik observed {\em critical phenomena} close to the critical parameter $\eta_\star$ that separates subcritical from supercritical data, and hence marks the threshold of black hole formation.  Specifically, Choptuik noted that the black hole mass found in supercritical evolutions scales with 
\begin{equation} \label{mass_scaling}
    M_{\rm BH} \simeq (\eta - \eta_\star)^\gamma,
\end{equation}
where $\gamma$ is the {\em critical exponent}, and that, close to criticality, the initial data evolve to approach a {\em self-similar critical solution}.  Choptuik found the critical exponent and the critical solution to be {\em unique} in his simulations of massless scalar fields, i.e.~independent of the initial data or their parametrization.

Inspired by Choptuik's discovery, numerous authors have studied similar phenomena in the gravitational collapse of other matter models, for different asymptotics, or relaxing the assumption of spherical symmetry (see, e.g., \cite{GunM07} for a review).  This body of work has resulted in a heuristic understanding of critical phenomena, at least in the context of spherical symmetry.   The critical exponent and the critical solution depend on the matter model, but are unique for each matter model.  Depending on the matter model, the critical solution can be either {\em discretely self-similar}  (DSS, e.g.~scalar fields) or {\em continuously self-similar} (CSS, e.g.~perfect fluids, see \cite{EvaC94} for an example).  The scaling law (\ref{mass_scaling}) can then be understood from perturbations of the self-similar critical solution; in particular, the critical exponent $\gamma$ is given by the inverse of the Lyapunov exponent of those perturbations (see, e.g., \cite{KoiHA95,Mai96}).  As pointed out by \cite{GarD98}, similar scaling applies to all dimensional, global quantities characterizing the evolution; based on dimensional arguments, the maximum energy density encountered in subcritical evolutions, for example, scales with
\begin{equation} \label{rho_scaling}
    \rho_{\rm max} \simeq (\eta_\star - \eta)^{- 2\gamma}
\end{equation}
(we have assumed in both Eqs.~\ref{mass_scaling} and \ref{rho_scaling} that $\eta > \eta_\star$ corresponds to supercritical data).  It has also been noted that, for matter models that display a DSS critical solution, the oscillations in the critical solution are reflected by a periodic ``wiggle" that is superimposed on the scaling laws (\ref{mass_scaling}) and (\ref{rho_scaling}) (see \cite{Gun97,HodP97}).

The situation is less clear in the absence of spherical symmetry.  Studying critical collapse of scalar fields, for example, the authors of \cite{ChoHLP03b} found that aspherical deformations may trigger an aspherical unstable mode that leads to a ``bifurcation" of the critical solution -- a result that was later confirmed by \cite{Bau18}.  Note that the existence of this instability does not seem to contradict \cite{MarG99}, who used a linear perturbation analysis to show that all nonspherical modes are stable, since the instability appears only for large deformations, well in the nonlinear regime (see \cite{Bau18}).  For sufficiently stiff ultrarelativistic fluids, unstable aspherical modes exist even in the linear regime (see \cite{Gun02,CelB18}).

For the above examples of scalar fields and fluids there exists a spherically symmetric critical solution, so that there is at least a limit in which the critical behavior is well understood.  This is no longer the case for matter models that do not allow spherically symmetric solutions.  The most important example is the critical collapse of gravitational waves in the absence of any matter, which we expect to display properties of gravity alone.  While critical phenomena in this vacuum collapse were first reported by \cite{AbrE93,AbrE94}, it has been very difficult to reproduce these results (see, e.g., Table I in \cite{HilBWDBMM13} for a summary of various different attempts). Significant progress was made by \cite{HilWB17}, who evolved so-called Brill wave initial data (see \cite{Bri59}), found a critical exponent similar to the value of $\gamma \simeq 0.37$ reported by \cite{AbrE93}, but found no convincing evidence of self-similarity.  Moreover, \cite{HilWB17} found a bifurcation, reminiscent of that reported by \cite{ChoHLP03b,Bau18}, with two separate black holes forming away from the center.  Quite recently, this result was confirmed by \cite{LedK21} who considered both Brill and (non-linear) Teukolsky waves (see \cite{Teu82}) as initial data.  Moreover, \cite{LedK21} found that these different initial data sets lead to different behavior near the black hole threshold (in agreement with the discussion of \cite{HilBWDBMM13}) and report different critical exponents for the different families of initial data.  The authors of \cite{LedK21} also report that they do not observe a universal self-similar solution in the limit of criticality.  All of this suggests the absence of a universal, strictly self-similar critical solution for the collapse of vacuum gravitational waves.  

Suspecting that properties of critical phenomena in the collapse of gravitational waves are related to the absence of a spherically symmetric critical solution, \cite{BauGH19} (hereafter BGH) studied critical collapse of electromagnetic waves.  Electromagnetic waves share with gravitational waves the absence of spherically symmetric solutions, but they share with scalar fields a very similar form of the evolution equations.  Since experience shows that the latter are easier to handle numerically than the former, electromagnetic waves provide a useful framework for exploring critical phenomena in the absence of spherical symmetry.  Focusing on dipole waves, BGH found an approximately DSS critical solution, but reported that this self-similarity is not exact.  Moreover, despite the restriction to dipole waves, BGH found that this critical solution can at best be approximately universal.  

In this paper we generalize the results of BGH and study gravitational collapse of electromagnetic waves with different multipole moments.  We argue that the symmetry of different multipole solutions alone rules out the existence of a unique critical solution, at least globally.  We then perform numerical simulations to fine-tune families of dipole and quadrupole data to the onset of black hole formation.  Unlike the dipole families previously considered by BGH, we find that the quadrupole data result in a bifurcation very similar to that reported by \cite{HilWB17,LedK21} for gravitational waves, with two centers of collapse forming on the axis but away from the center -- confirming our expectation that the critical solution cannot be unique.  Because of this bifurcation, it is significantly harder to analyze the properties of the quadrupole solutions than those of dipole solutions, both numerically and conceptually.  Accordingly, some of our results are of a qualitative rather than quantitative nature, but we nevertheless believe that our study provides interesting and important insights into the effects of multipoles on critical phenomena in gravitational collapse.

Our paper is organized as follows.  In Section \ref{sec:EM} we review Maxwell's equations, and provide analytical solutions describing electromagnetic waves in flat Minkowski spacetimes.  In Section \ref{sec:numerics} we describe our numerical simulations, starting with initial data based on the analytical solutions of Section \ref{sec:EMwaves}.  We present our numerical results in \ref{sec:results}, and close with a summary and discussion in Section \ref{sec:summary}.  Throughout this paper we adopt geometrized units with $G = 1 = c$.

\section{Electrodynamics}
\label{sec:EM}

\subsection{3+1 decomposition of spacetime}
\label{sec:3+1}

In our calculations we adopt a ``3+1" decomposition of spacetime and write the line element as
\begin{equation}
    ds^2 = g_{ab} dx^a dx^b = - \alpha^2 dt^2 + 
    \gamma_{ij} (dx^i + \beta^i dt)(dx^j + \beta^j dt).
\end{equation}
Here $g_{ab}$ is the spacetime metric, $\alpha$ the lapse function, $\gamma_{ij}$ the spatial metric induced on spatial slices, and $\beta^i$ the shift vector.  We adopt the convention that indices $a, b, \ldots$ run over spacetime components, while indices $i, j, \ldots$ run over spatial components only.  In terms of the lapse and the shift, the unit vector $n^a$ normal on the spatial slices can be written as
\begin{equation}
    n_a = (-\alpha,0,0,0),~~~~~~n^a = \alpha^{-1} (1, - \beta^i).
\end{equation}
The mean curvature $K = \gamma^{ij} K_{ij}$, i.e.~the trace of the extrinsic curvature $K_{ij}$, can be written as the negative divergence of the normal vector,
\begin{equation}
    K = - \nabla_a n^a,
\end{equation}
where $\nabla_a$ denotes the covariant derivative associated with the spacetime metric $g_{ab}$.

\subsection{Maxwell's equations}
\label{sec:maxwell}

We express Maxwell's equations in terms of a vector potential $A_a$, so that the Faraday tensor can be written as 
\begin{equation} \label{faraday}
    F_{ab} = \nabla_a A_b - \nabla_b A_a. 
\end{equation} 
In terms of the Faraday tensor, the stress-energy tensor of the electromagnetic fields is given by
\begin{equation}
    T^{ab} = \frac{1}{4 \pi} \left( F^{ac} F^{b}{}_c - \frac{1}{4} \, g^{ab} F_{cd} F^{cd} \right).
\end{equation}

Without loss of generality we may choose an electromagnetic gauge in which $\Phi \equiv n^a A_a = 0$, so that $A_a$ becomes purely spatial.  In the absence of charges, Maxwell's equations may then be written as
\begin{subequations}\label{maxwell}
\begin{align} 
    d_t A_i & = - \alpha E_i  \label{Adot}\\
    d_t E^i & = - D_j (\alpha D^j A^i) + D_j (\alpha D^i A^j) + \alpha K E^i, \label{Edot}
\end{align}
\end{subequations}
together with the Gaussian constraint 
\begin{equation}
    D_i E^i = 0.
\end{equation}
Here $E^a = F^{ab} n_b$ is the electric field as observed by a normal observer, $D_i$ the covariant derivative associated with the spatial metric $\gamma_{ij}$,  and $d_t \equiv \partial_t - {\mathcal L}_\beta$, where ${\mathcal L}_\beta$ denotes the Lie derivative along $\beta^i$.  In terms of these quantities we may rewrite the Faraday tensor (\ref{faraday}) as
\begin{equation}
    F_{ab} = D_a A_b - D_b A_a + n_a E_b - n_b E_a.
\end{equation}
The magnetic field as observed by a normal observer is given by
\begin{equation} 
    B^a = \frac{1}{2} \, \epsilon^{abcd} n_b F_{dc},
\end{equation}
where $\epsilon^{abcd}$ is the spacetime Levi-Civita tensor, or the more familiar expression
\begin{equation} \label{B}
    B^i = \epsilon^{ijk} D_j A_k,
\end{equation}
where $\epsilon^{abc} \equiv n_d \epsilon^{dabc}$ is the spatial Levi-Civita tensor.  Note that both $E^a$ and $B^a$ are purely spatial, $n_a E^a = 0$ and $n_a B^a = 0$.

We compute the energy density $\rho$ as measured by a normal observer from
\begin{equation} \label{rho}
    \rho \equiv n_a n_b T^{ab} = \frac{1}{8 \pi} \left( E_i E^i + B_i B^i \right),
\end{equation}
and the momentum density, i.e.~the Poynting vector, from
\begin{equation} \label{Poynting}
    S^i \equiv - \gamma^{ia} n^b T_{ab} = \frac{1}{4 \pi} \epsilon^{ijk} E_j B_k.
\end{equation}

Throughout this paper we will assume axisymmetry, which, in adapted coordinates, is generated by a Killing vector field $\xi^a = \partial / \partial \varphi$.  In twist-free axisymmetry (see \cite{Ger71}), we can then reduce Maxwell's equations (\ref{maxwell}) to a single wave equation for $A_\varphi \equiv \xi^a A_a$ and its conjugate variable $E^\varphi$.

All solutions that we discuss are also either symmetric or antisymmetric across the equatorial plane, which singles out a well-defined central observer.  We note that the density $\rho$ in (\ref{rho}) depends on the slicing of the spacetime (but not on the spatial coordinates) except at the center, where the central observer represents a preferred normal observer.

\subsection{Electromagnetic waves in flat spacetimes}
\label{sec:EMwaves}

In the absence of gravity, i.e.~in flat spacetimes, we may adopt the Minkowski metric in Maxwell's equations~(\ref{maxwell}), so that $\alpha = 1$, $\beta^i = 0$ and $K = 0$, and so that all covariant derivatives reduce to their usual flat expressions (in Cartesian coordinates, in particular, they reduce to partial derivatives).  We may then derive regular analytical solutions to Maxwell's equations, representing electromagnetic waves of different multipole moments $\ell$, as discussed in Appendix \ref{sec:ana_sols}.  In the following we list results for dipole, quadrupole, and octupole waves that feature a moment of time symmetry at $t =0$. 

\subsubsection{Dipole waves}
\label{sec:dipole}

In spherical polar coordinates, an analytical dipole solution, i.e.~for $\ell = 1$, is given by $A^{\hat r} = A^{\hat \theta} = 0$ and
\begin{equation}\label{DipoleSolution}
    A^{\hat \varphi} = \mathcal{A}\sin{\theta}\left(\frac{e^{-u^2} - e^{-v^2}}{(r/\sigma)^2} + \frac{2u e^{-u^2} - 2v e^{-v^2}}{r/\sigma} \right),
\end{equation}
where $\mathcal{A}$ is a dimensionless amplitude, $\sigma$ a constant with units of length, and we have introduced the dimensionless abbreviations
\begin{equation} \label{uv}
    u = \frac{r - t}{\sigma},~~~~ v = \frac{r + t}{\sigma}.
\end{equation}
Note also that we have expressed (\ref{DipoleSolution}) in terms of an orthonormal vector component, denoted by the ``hat"; the corresponding orthonormal basis vector is ${\bf e}_{\hat \varphi} = (r \sin\theta)^{-1} \, \partial / \partial \varphi$.    Here and in the following physical units enter through the constant $\sigma$ only, and we will therefore express all dimensional results in units of $\sigma$.

We can compute the electric field $E^i$ corresponding to the solution (\ref{DipoleSolution}) from (\ref{Adot}); evaluating the result at the initial time $t = 0$ yields
\begin{equation}\label{DipoleEField}
    E^{\hat \varphi} = - 8 \mathcal{A}\frac{r\sin{\theta}}{\sigma^2}e^{-(r / \sigma)^2} ~~~~~~~~~~(t = 0).
\end{equation}
We can similarly compute the magnetic field $B^i$ for the solution (\ref{DipoleSolution}) from (\ref{B}).  Expanding $A^i$ and $E^i$ about the center shows that, to leading order, both are linear in $r$ there.  The magnetic field of the dipole wave (\ref{DipoleSolution}), however, takes a non-zero value at the center.  As a result, the energy density (\ref{rho}) of the dipole wave (\ref{DipoleSolution}) also does not vanish at the center,
\begin{equation}\label{Dipolerhocenter}
\rho = \frac{32 \mathcal{A}^2}{9 \pi} \,\frac{t^2 (3 \sigma^2 - 2 t^2)^2}{\sigma^8} \, e^{-2 (t/\sigma)^2} ~~~~~~(r = 0).
\end{equation}
In fact, in our numerical simulations of dipole waves, even when coupled to gravity, we encounter the largest densities at the center (see Fig.~\ref{fig:rho} below). 

Finally, note that $A^i$ and $E^i$ for the dipole solution are {\em symmetric} across the equator, as can be seen from Eqs.~(\ref{DipoleSolution}) and (\ref{DipoleEField}) above.

\subsubsection{Quadrupole waves}
\label{sec:quadrupole}

An analytical $\ell = 2$ quadrupole solution to Maxwell's equations (\ref{maxwell}) in flat spacetimes is given by
\begin{equation}\label{QuadrupoleSolution}
\begin{split}
     A^{\hat \varphi} = & 
     \mathcal{A}\sin{\theta}\cos{\theta}\Biggl\{\frac{e^{-u^2} - e^{-v^2}}{(r/\sigma)^3} + \frac{2u e^{-u^2} - 2v e^{-v^2}}{(r/\sigma)^2} \\ &+
     \frac{4u^2 e^{-u^2} -4v^2e^{-v^2} -2e^{-u^2} +2 e^{-v^2}}{3 r / \sigma } \Biggr\},
\end{split}
\end{equation}
We again compute the electric field from (\ref{Adot}) to find, at the initial time $t=0$,
\begin{equation}\label{QuadrupoleEField}
    E^{\hat \varphi}= -\frac{16\mathcal{A}}{3}\,\frac{r^2\sin{\theta}\cos{\theta}}{\sigma^3}e^{-(r/\sigma)^2} ~~~~~~~~~~(t = 0).
\end{equation}
Expanding $A^i$ and $E^i$ about the center shows that, to leading order, the quadrupole fields are now quadratic in $r$, while the magnetic field, computed from (\ref{B}), is now linear in $r$.  Accordingly, the energy density of the quadrupole wave (\ref{QuadrupoleSolution}) vanishes identically at the center.  This is consistent with results from our numerical simulations of quadrupole waves, even when they are coupled to gravity, where we encounter the maximum densities on the symmetry axis, but away from the center (see Fig.~\ref{fig:rho} below).

Note also that $A^i$ and $E^i$ for the quadrupole solution (\ref{QuadrupoleSolution}) and (\ref{QuadrupoleEField}) are {\em antisymmetric} across the equator, unlike the dipole solution (\ref{DipoleSolution}), which was symmetric (see Fig.~\ref{fig:Axi} below). Since these symmetries are maintained even when the solutions are coupled to gravity,  as we verified numerically,\footnote {Note that the electromagnetic fields enter the stress-energy tensor quadratically, so that the sources for the gravitational fields are symmetric for either symmetric or antisymmetric electromagnetic fields.}
this finding alone indicates that the critical solution for quadrupole waves cannot be the same as that for dipole waves.  This argument alone demonstrates that the critical solution for the gravitational collapse of electromagnetic waves cannot be unique, at least not globally.

\subsubsection{Octupole waves}
\label{sec:octupole}

While we will focus on dipole and quadrupole waves in our numerical simulations, we briefly discuss an $\ell = 3$ octupole solution to Maxwell's equations (\ref{maxwell}) in flat spacetimes, 
\begin{equation}\label{OctupoleSolution}
\begin{split}
     A^{\hat \varphi} = & \mathcal{A}\left(5\cos^2{\theta}-1 \right)\sin{\theta}\Biggl\{\frac{e^{-u^2} - e^{-v^2}}{(r/\sigma)^4} \\
     & + \frac{2u e^{-u^2} - 2v e^{-v^2}}{(r/\sigma)^3} \\ &+
     \frac{8u^2 e^{-u^2} -8v^2e^{-v^2} -4e^{-u^2}+ 4 e^{-v^2}}{5 (r/\sigma)^2} \\ 
    & +
    \frac{8u^3 e^{-u^2} -8v^3e^{-v^2} -12u e^{-u^2}+12v e^{-v^2}}{15 r / \sigma} \Biggr\},
\end{split}
\end{equation}
 in order to highlight some qualitative difference from both the dipole and the quadrupole data.  As before we compute the electric field from (\ref{Adot}); evaluating the result for the initial time $t=0$ yields
\begin{equation}\label{OctupoleEField}
    E^{\hat \varphi}= -\frac{32\mathcal{A}}{15} \,\frac{r^3(\cos{\theta}^2-1)\sin{\theta}}{\sigma^5}e^{-(r/\sigma)^2}
    ~~~~~(t = 0).
\end{equation}
Expanding the fields about the center shows that $A^i$ and $E^i$ now scale with $r^3$ there, and $B^i$ with $r^2$, so that the energy density again vanishes at the center. Note also that octupole waves are again {\em symmetric} across the equator. We see that octupole waves differ qualitatively from both dipole waves (in terms of the location of the maximum densities) and quadrupole waves (in terms of the symmetry). Therefore, we also expect the corresponding critical solutions for octupole waves to be different from both dipole and quadrupole waves -- again at least globally.

\section{Numerics}
\label{sec:numerics}

While we can describe electromagnetic waves in flat spacetimes analytically, this is no longer possible, of course, in curved spacetimes, when we take into account the self-gravity of the electromagnetic radiation.  Instead, we construct such solutions to the Einstein-Maxwell system numerically, adopting the approach described in this Section.

\subsection{Initial Data}
\label{sec:indata}

We construct initial data that are time symmetric (i.e.~$K_{ij} = 0$) and conformally flat (i.e.~$\gamma_{ij} = \psi^4 \eta_{ij}$, where $\psi$ is the conformal factor and $\eta_{ij}$ the flat metric).  As our initial data for the electromagnetic fields we adopt expressions that reduce to those of Section \ref{sec:EMwaves}, evaluated at $t=0$, in the limit of weak fields.  Specifically, we choose $A^i = 0$ initially, so that, according to (\ref{B}), $B^i = 0$ also.  This means that the momentum density (\ref{Poynting}) of the electromagnetic fields vanishes initially, and that the momentum constraint is satisfied identically.

This leaves us with having to solve the Hamiltonian constraint 
\begin{equation} \label{hamilton}
    \bar D^2 \psi = - 2 \pi \psi^5 \rho
\end{equation}
only, where $\bar D^2$ is the flat Laplace operator and $\rho$ the energy density (\ref{rho}).  We solve this equation iteratively as follows.  In order to help with the convergence of this iteration, we adopt as the initial electric fields not the expressions (\ref{DipoleEField}), (\ref{QuadrupoleEField}) or (\ref{OctupoleEField}) themselves, but rather those expressions divided by $\psi^6$ (see also BGH).  In practice, we start with an initial guess for $\psi$, then compute the electric field given our choice of the amplitude $\mathcal{A}$, evaluate the density $\rho$ from (\ref{rho}), and then solve the Hamiltonian constraint (\ref{hamilton}) for a new conformal factor $\psi$.  We repeat the process until convergence to within a desired tolerance has been achieved.  For weak electromagnetic fields we have $\psi \rightarrow 1$, so that our numerical solutions approach the analytical solutions of Section \ref{sec:EMwaves} in this regime.

In the absence of gravity, electrodynamics is linear, which allowed the
identification of well-defined multipole moments in Section \ref{sec:EMwaves}.  In the context of general relativity, however, different multipole moments will couple to each other through the nonlinearities in Einstein's equations.  Since Einstein's equations preserve the symmetry across the equator, we expect that modes of odd (even) $\ell$ will be coupled to other modes of odd (even) $\ell$ only.  In the following we will still refer to ``dipole" and ``quadrupole" waves, expecting that our data will be dominated by the corresponding multipole, but understanding that nonlinear coupling introduces other multipoles as well.

\subsection{Evolution}
\label{sec:evolve}

We evolve our initial data using a numerical code that implements the Baumgarte-Shapiro-Shibata-Nakamura (BSSN) formalism \cite{NakOK87,ShiN95,BauS99} in spherical polar coordinates.  Details of our numerical approach are described in \cite{BauMCM13,BauMM15}; in particular, we use a reference-metric formalism (see, e.g., \cite{ShiUF04,BonGGN04,Bro09,Gou12}) together with an appropriate rescaling of all tensorial variables to handle the coordinate singularities at the origin and on the axis analytically.  All spatial derivatives are evaluated using a fourth-order finite-difference method.  The latest version of our code, which we have also used in BGH, adopts a fourth-order Runge-Kutta time integrator rather than the ``partially implicit Runge-Kutta" method described in \cite{BauMCM13} (see, e.g., Fig.~3 in BGH for a demonstration of fourth-order convergence).

As discussed in BGH, we evolve the electromagnetic fields in terms of rescaled variables $a_\varphi \equiv  A_\varphi/(r \sin \theta)$ and $e^\varphi \equiv r \sin \theta \, E^\varphi$.

A new feature in our simulations here concerns the allocation of the radial grid points.  Following \cite{RucEB18}, the radial grid is constructed by mapping a uniform grid in a variable $x$, covering the interval $[0,1]$, to our radial variable $r = r(x)$, covering the interval $[0, r_{\rm max}]$.  We now adopt the function 
\begin{equation} \label{map}
    r = \frac{r_{\rm max}}{1 + A} \left( \frac{\sinh(s_p x)}{\sinh s_p} + A \frac{\tanh(t_p x)}{\tanh t_p} \right)
\end{equation}
for this mapping, where $s_p$, $A$, and $t_p$ are dimensionless parameters.  For dipolar waves, which result in collapse at the origin, we choose $A = 0$ and $s_p = 6.57$, resulting in the same ``sinh" grid setup as used in BGH: it allows for a high, nearly uniform resolution near the origin, but an increasingly coarse, approximately logarithmic resolution at large separations from the origin.  For higher multipole moments, for which we observe collapse away from the origin, this resulted in unnecessarily high resolution near the origin, and hence an unnecessarily short timestep.  We therefore added the ``tanh" term in (\ref{map}), which makes it possible to construct a grid that is relatively coarse at the origin, becomes finer at some distance from the origin, but then becomes approximately logarithmic again at large separations.  For our simulations of the quadrupole waves we adopted $s_p = 6$, $A = 0.0015$, and $t_p = 50$. All results shown in Section \ref{sec:results} for the quadrupole waves were performed with $N_r = 256$ radial grid points and the outer boundary at $r_{\rm max} = 128$ (in units of $\sigma$), with $N_\theta = 48$ angular grid points (covering one hemisphere), and with a Courant factor of 0.4.

We evolve the fields using the ``1+log" slicing condition
\begin{equation}
    (\partial_t - \beta^i \partial_i) \, \alpha = - 2 \alpha K
\end{equation}
(see \cite{BonMSS95}), starting with the ``pre-collapsed" lapse $\alpha = \psi^{-2}$ as initial data.  We note that, in the simulations of \cite{CelB18,Bau18}, the 1+log slicing condition resulted in spatial slices that reflect the symmetry of the self-similar critical solutions.  On such preferred slices, slicing-dependent quantities take on invariant meanings; in the following we will therefore assume that the density $\rho$, defined in (\ref{rho}), provides an adequate diagnostic of our simulations.  As discussed in BGH, the ``Gamma-driver" shift condition did not allow us to obtain stable evolution close to the onset of black hole formation.  Using zero shift, however, we were able to complete subcritical solutions close to the black hole threshold.  In all simulations presented here we will therefore use zero shift, and will focus on subcritical solutions only.

\section{Results}
\label{sec:results}

\subsection{Minimum lapse and maximum density}
\label{sec:lapse_dens}

\begin{figure}[t]
    \centering
    \includegraphics[width = 0.5 \textwidth]{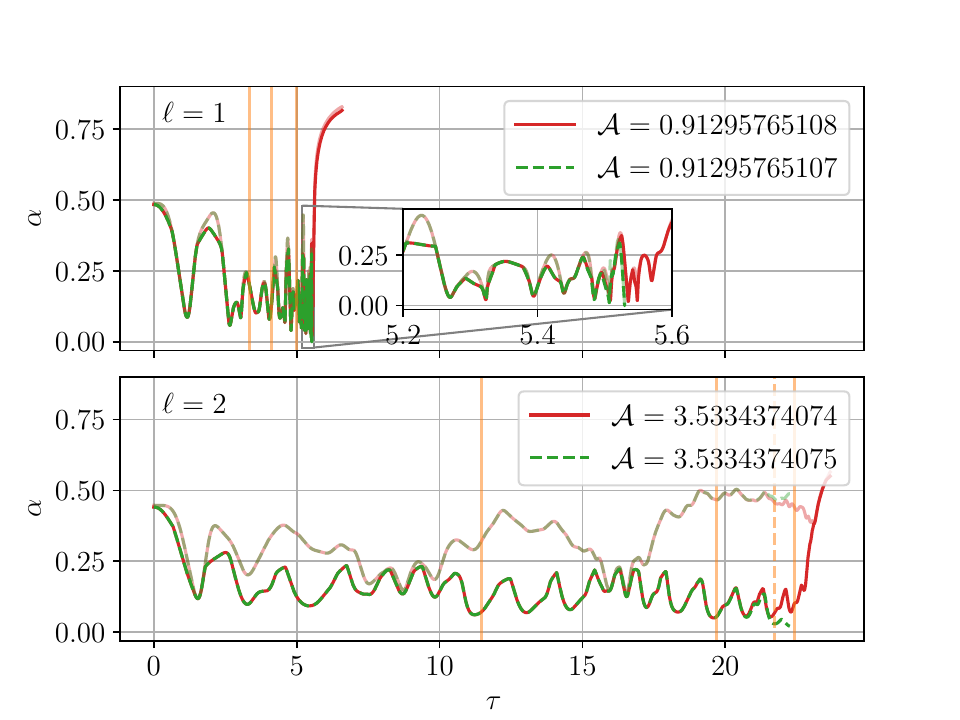}
    \caption{The lapse function $\alpha$ as a function of proper time $\tau$ as observed by an observer at the center, for dipole waves ($\ell = 1$) in the top panel and quadrupole waves ($\ell = 2$) in the bottom panel.  The dark lines represent the minimum values of the lapse on spatial slices with the same coordinate time as that of the central observer, while the faint lines represent values of the lapse at the center, both for subcritical solutions (the solid red lines) and supercritical solutions (the dashed green lines).  Note that, for most of the evolution, the minimum values of the lapse are found at the center for the dipole waves, but away from the center for quadrupole waves.  The solid vertical (orange) lines mark the times of the snapshots shown in Figs.~\ref{fig:Axi} through \ref{fig:rho};   Figs.~\ref{fig:rho_prop_R} and \ref{fig:rho_photon} include an additional snapshot at the time marked by the dashed vertical line.}
    \label{fig:lapse_of_t}.
\end{figure}

We start our analysis by bracketing the critical parameters $\mathcal{A}_\star$ for different multipoles.  In Fig.~\ref{fig:lapse_of_t} we show results for the lapse function $\alpha$ as a function of proper time $\tau$ for pairs of data bracketing the critical solution, for both dipole and quadrupole waves.  Here and in the following we refer to proper time as that as measured by an observer at the center.  The faint lines in the figure represent values of the lapse as measured by this central observer, while the dark lines represent minimum values of the lapse on spatial slices, i.e.~on slices with the same {\em coordinate} time as that of the central observer.

\begin{figure}[t]
    \centering
    \includegraphics[width = 0.5 \textwidth]{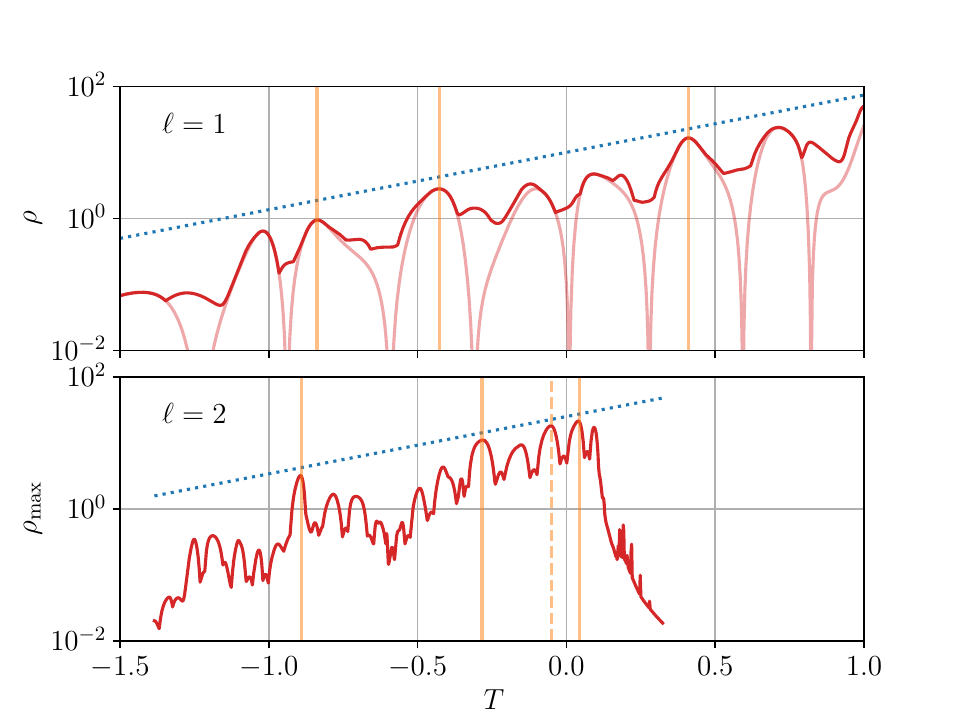}
    \caption{The density $\rho$ (see Eq.~\ref{rho}) as a function of the ``slow time" (\ref{slow_time}) for the subcritical solutions shown in Fig.~\ref{fig:lapse_of_t}.  We show results for dipole waves ($\ell = 1$) in the top panel and quadrupole waves ($\ell = 2$) in the bottom panel.  For dipole data we have included both the maximum values on a given slice of constant coordinate time (the dark lines) and values at the center (the faint lines) while, for quadrupole waves, we have included the former only, since the density vanishes identically at the center (see the discussion in Section \ref{sec:quadrupole}).  The dotted (blue) lines show the exponential growth $e^{2 T}$ expected for the density in a self-similar contraction, while the solid (dashed) vertical (orange) lines indicate the times of the snapshots shown in Figs.~\ref{fig:Axi} through \ref{fig:rho} (as well as \ref{fig:rho_prop_R} and \ref{fig:rho_photon}). }
    \label{fig:rho_of_T}
\end{figure}

\begin{figure*}[t]
    \centering
     \includegraphics[width = 0.45 \textwidth]{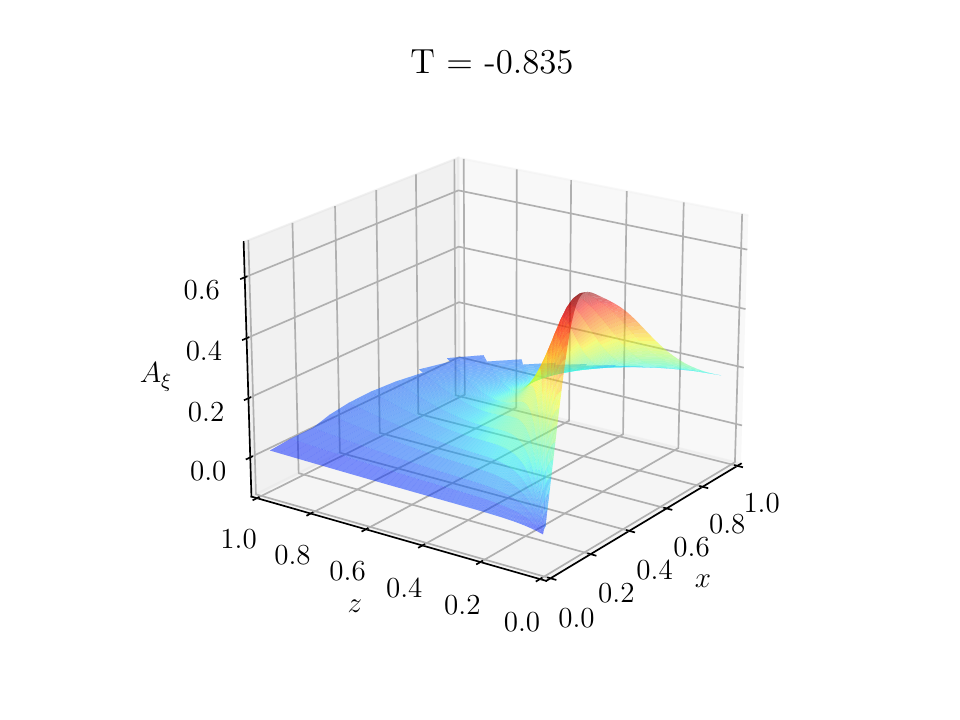}
     \includegraphics[width = 0.45 \textwidth]{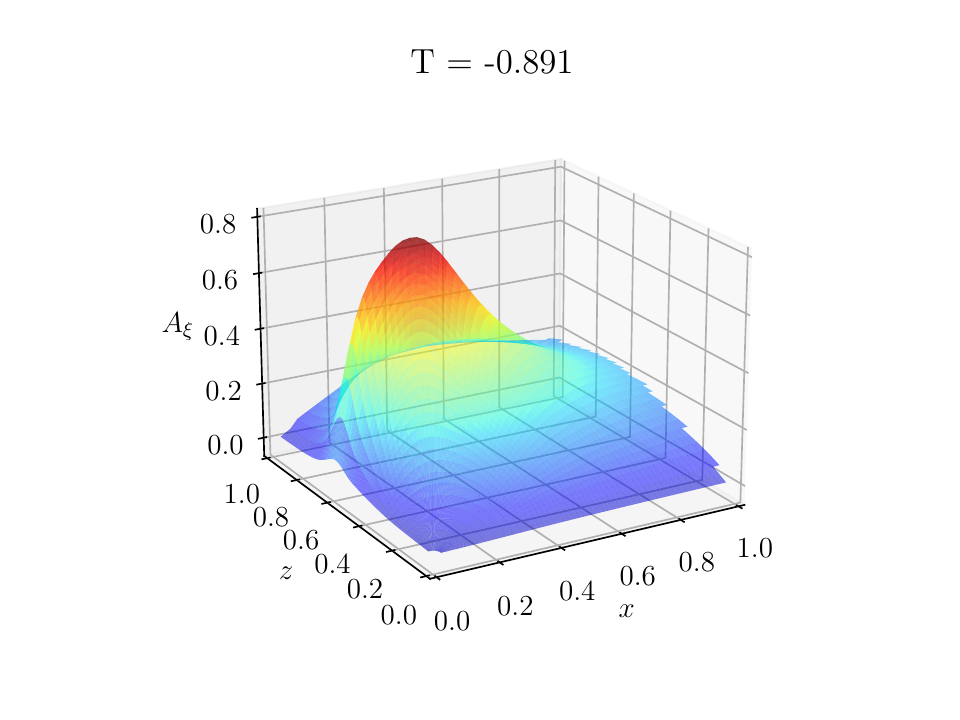}

     \includegraphics[width = 0.45 \textwidth]{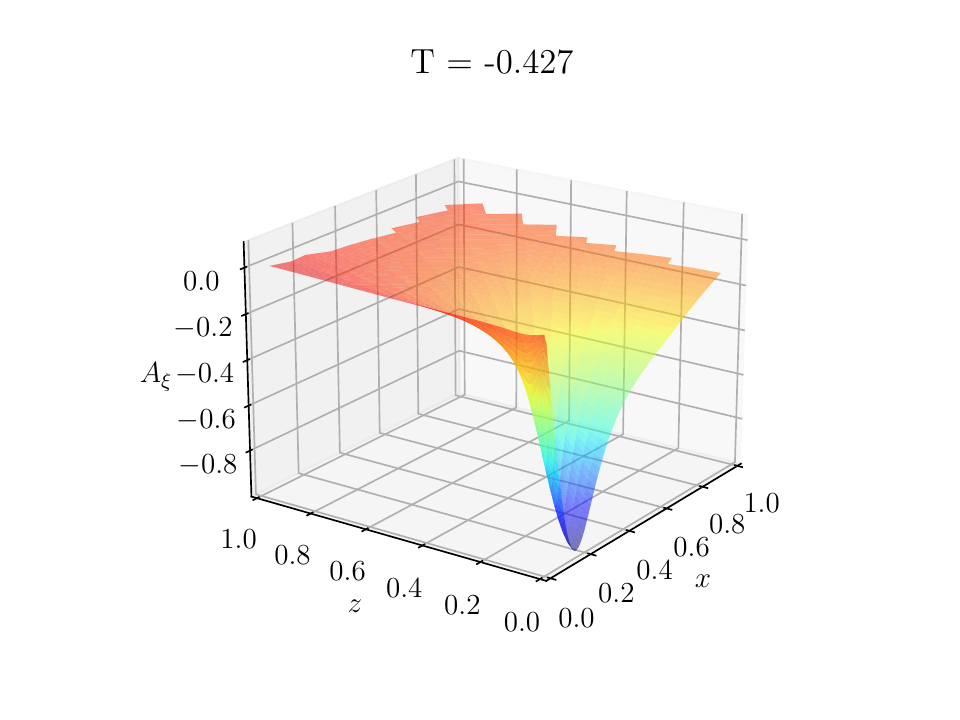}
     \includegraphics[width = 0.45 \textwidth]{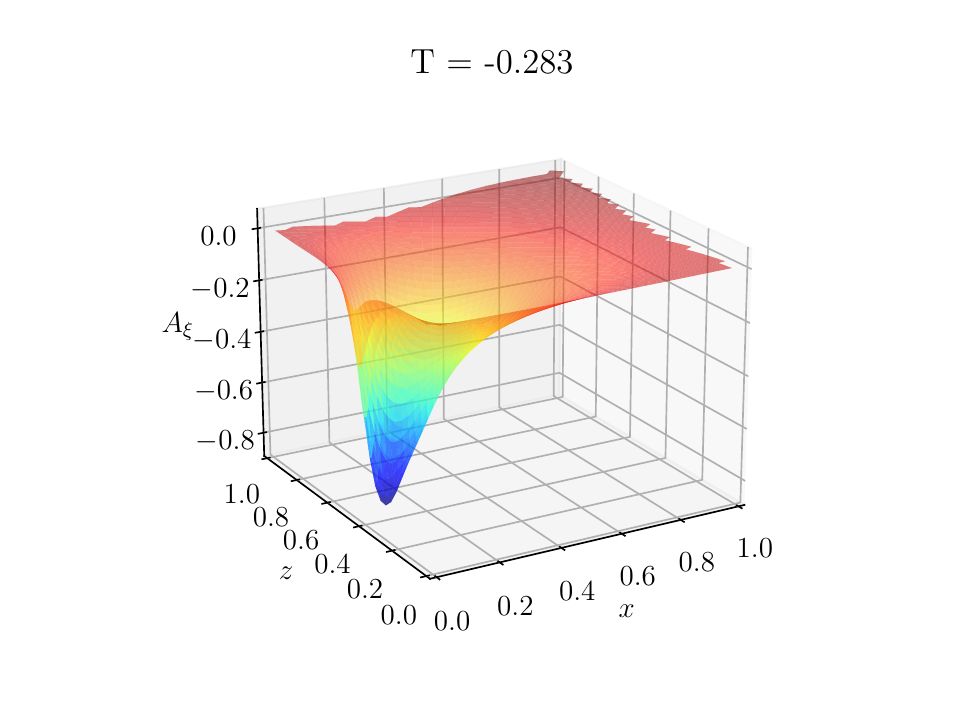}

     \includegraphics[width = 0.45 \textwidth]{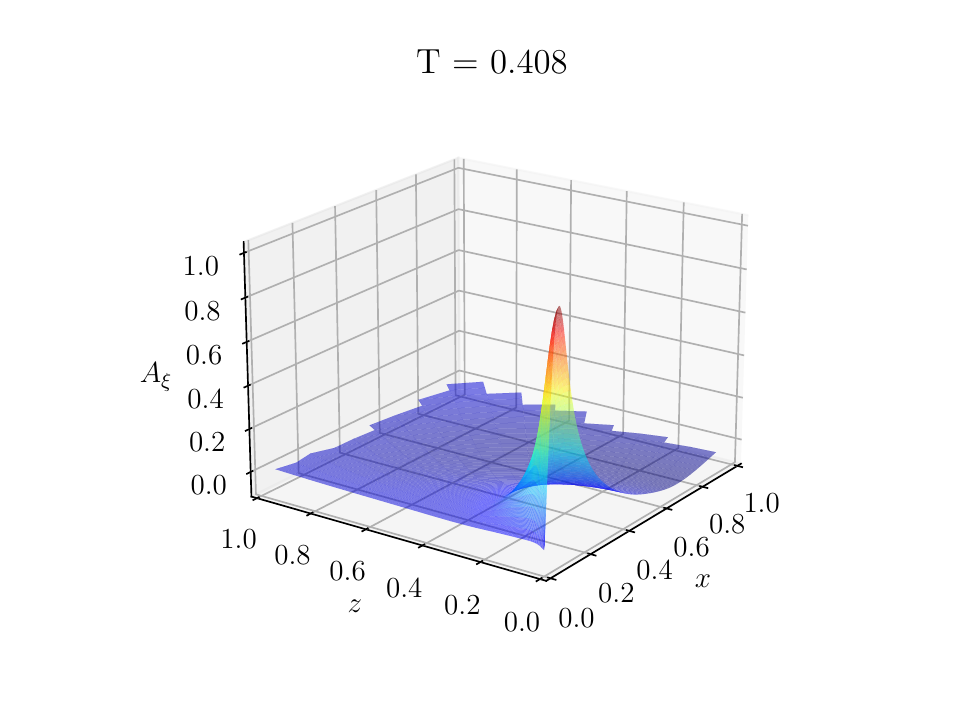}
     \includegraphics[width = 0.45 \textwidth]{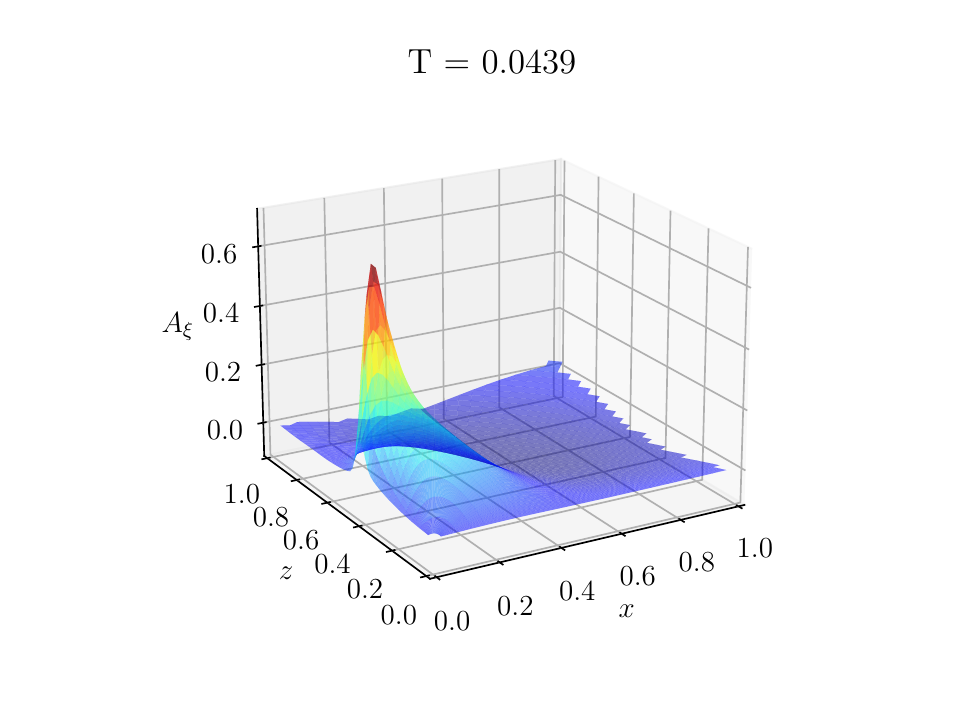}
    \caption{Snapshots of the vector potential $A_{\xi}$ (see Eq.~\ref{Axi}) for a near-critical evolution at the instants marked by the solid vertical lines in Figs.~\ref{fig:lapse_of_t} and \ref{fig:rho_of_T}.  We show results for dipole data in the left column, and quadrupole data in the right column.  Note that the dipole data are symmetric across the equator, while the quadrupole data are antisymmetric (see also the discussion in Sections \ref{sec:dipole} and \ref{sec:quadrupole}).  }
    \label{fig:Axi}
\end{figure*}

\begin{figure}[t]
    \centering
     \includegraphics[width = 0.4 \textwidth]{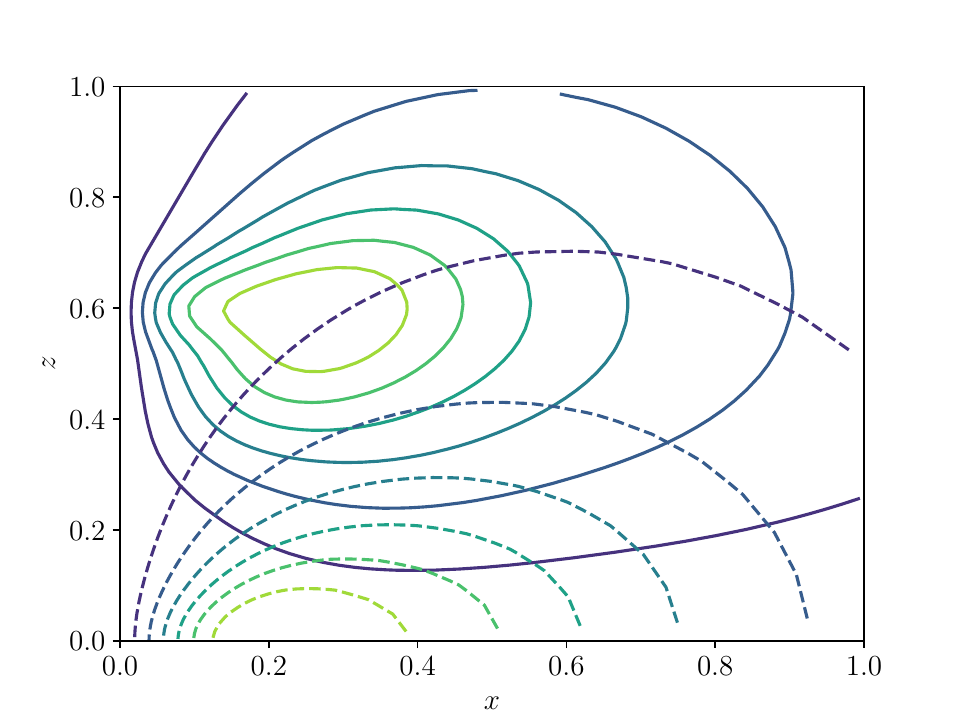}

     \includegraphics[width = 0.4 \textwidth]{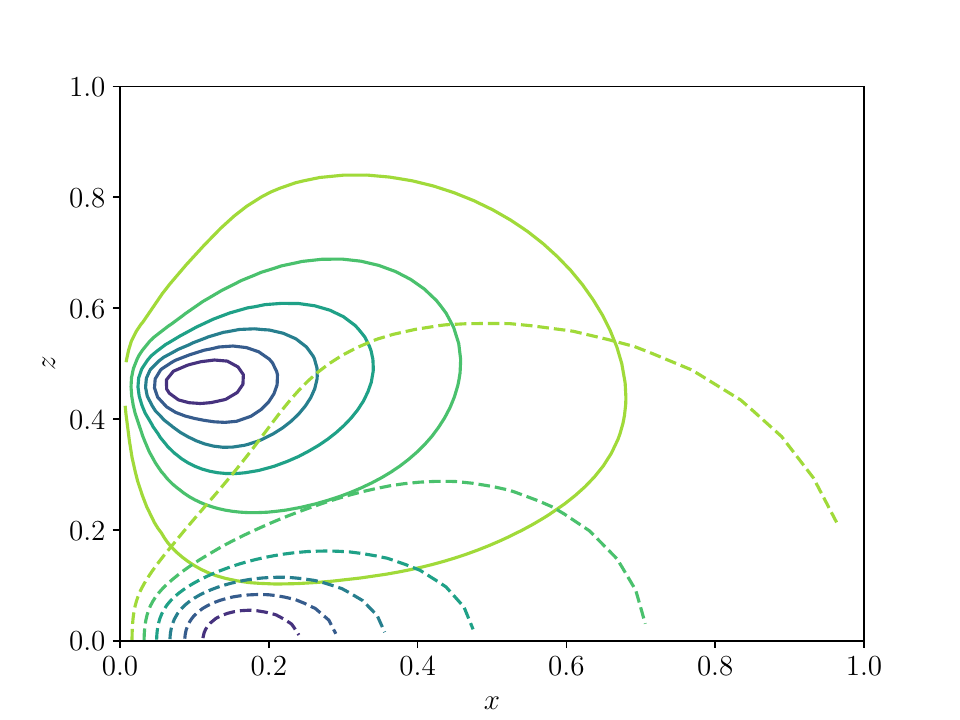}

     \includegraphics[width = 0.4 \textwidth]{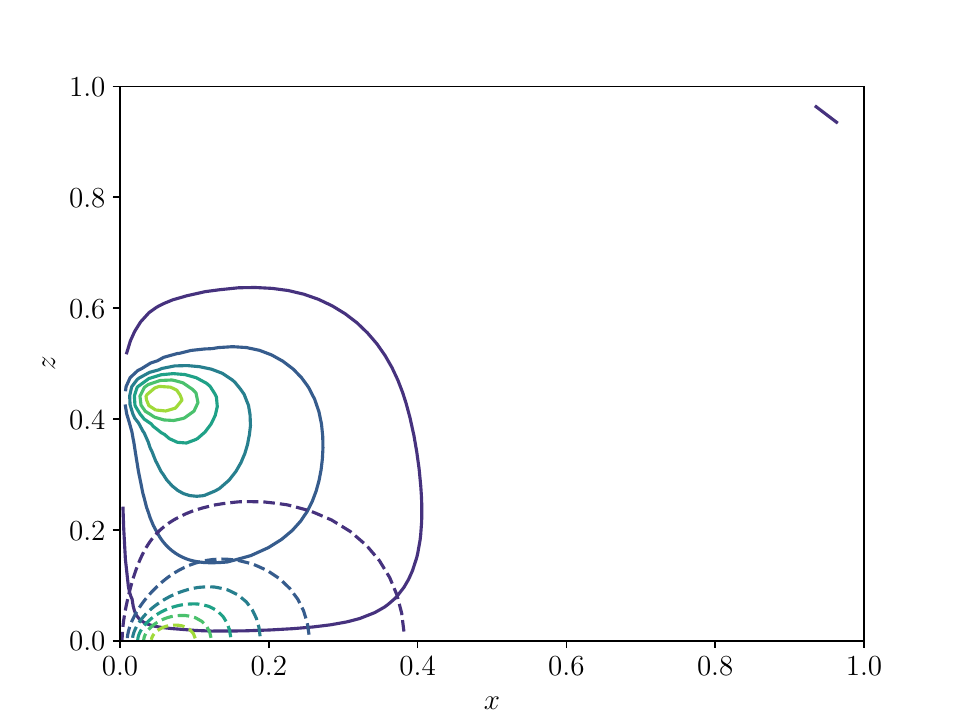}
    \caption{ Equidistant contours of the vector potential $A_\xi$ at the three different times shown in Fig.~\ref{fig:Axi}.  Here the top (middle, bottom) panel corresponds to the two times shown in the top (middle, bottom) row of Fig.~\ref{fig:Axi}.  Solid lines mark contours for the quadrupole data in the right column of Fig.~\ref{fig:Axi}, while dashed lines mark those for the dipole data in the left column.}
    \label{fig:Axi_contour}
\end{figure}

\begin{figure*}[t]
    \centering
     \includegraphics[width = 0.45 \textwidth]{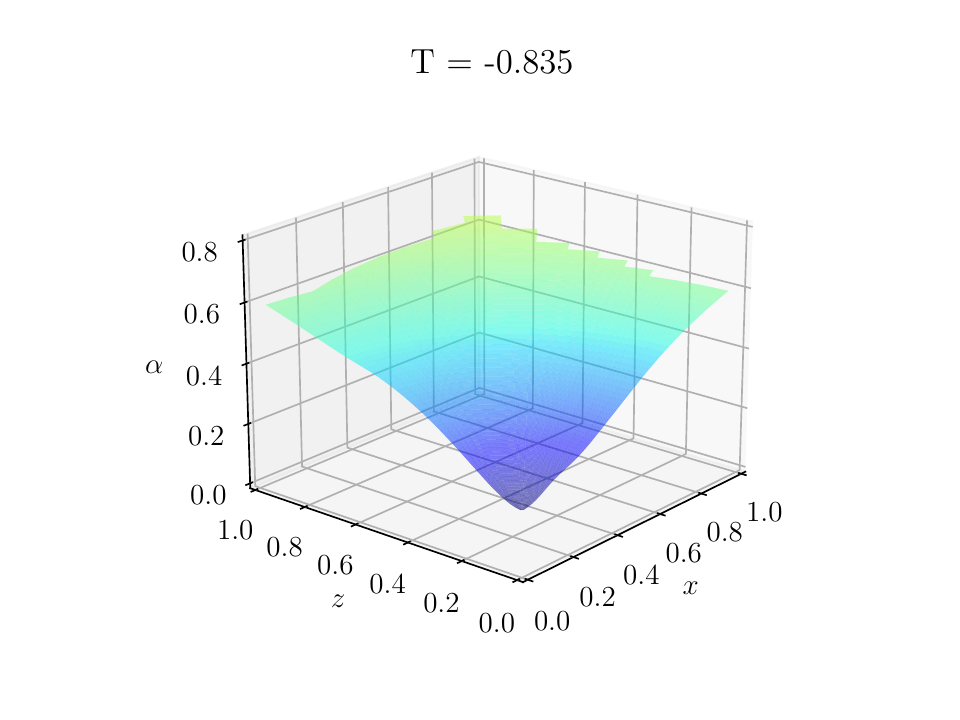}
     \includegraphics[width = 0.45 \textwidth]{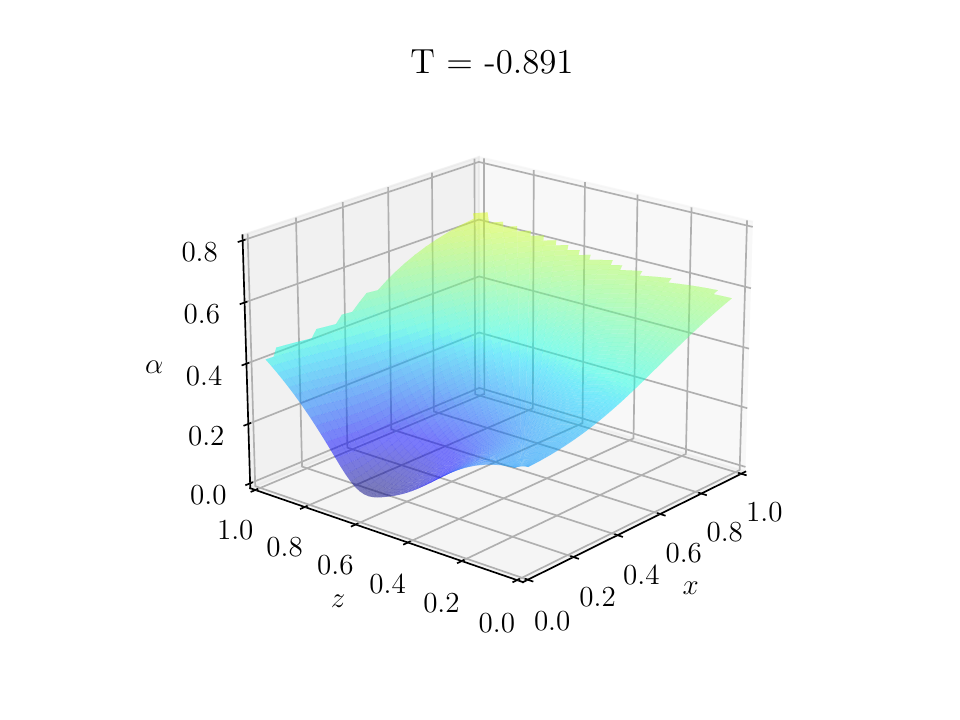}

     \includegraphics[width = 0.45 \textwidth]{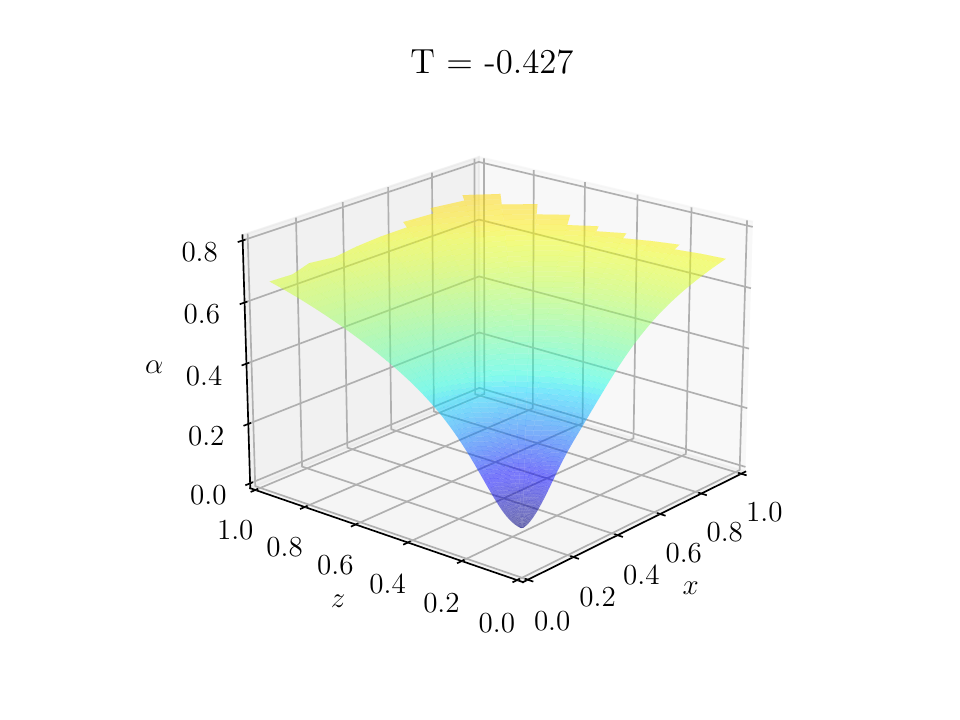}
     \includegraphics[width = 0.45 \textwidth]{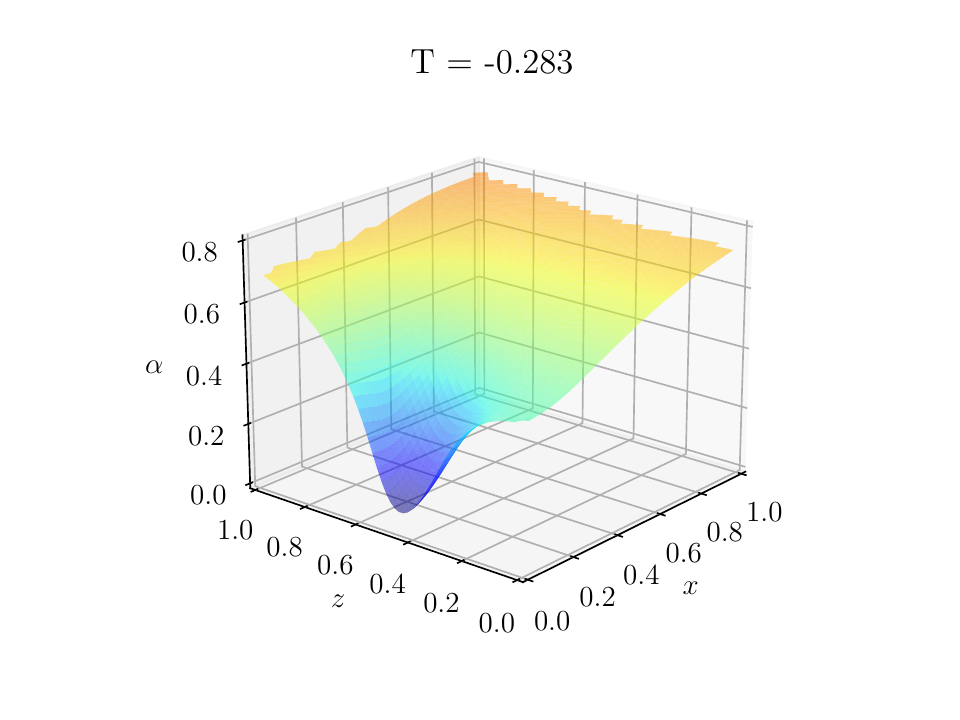}

     \includegraphics[width = 0.45 \textwidth]{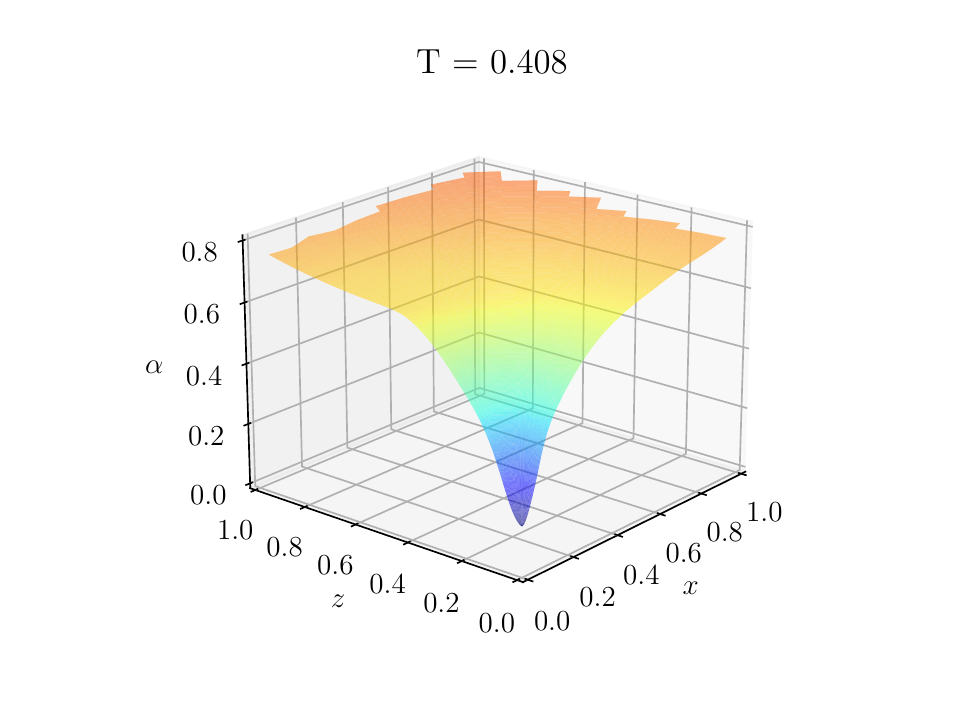}
     \includegraphics[width = 0.45 \textwidth]{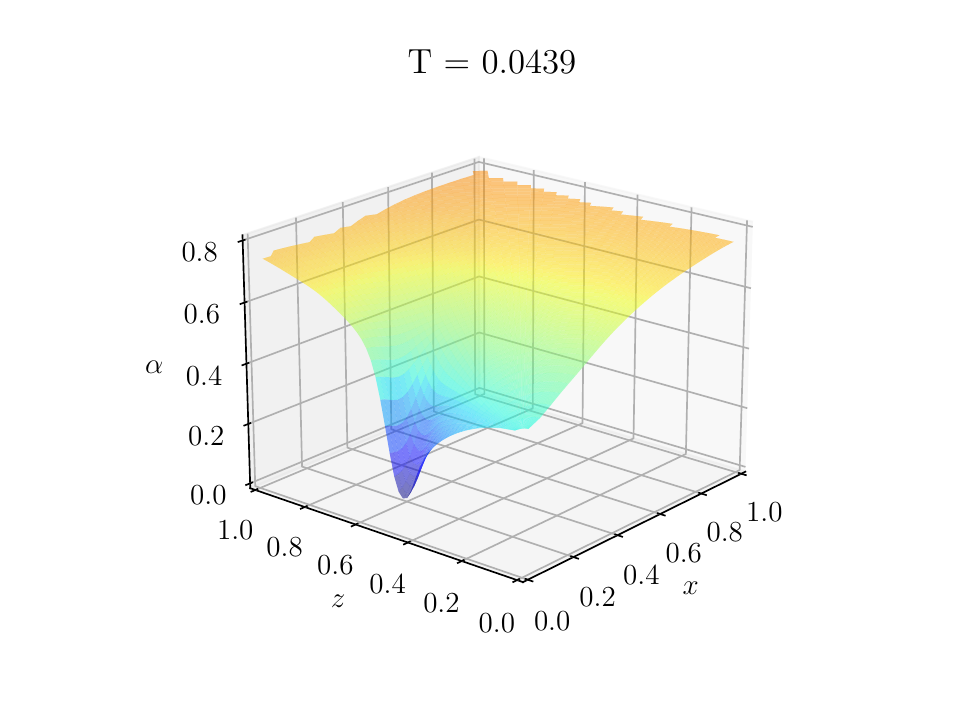}
     \caption{Same as Fig.~\ref{fig:Axi}, but for the lapse function $\alpha$.}
    \label{fig:lapse}
\end{figure*}

\begin{figure*}[t]
    \centering
     \includegraphics[width = 0.45 \textwidth]{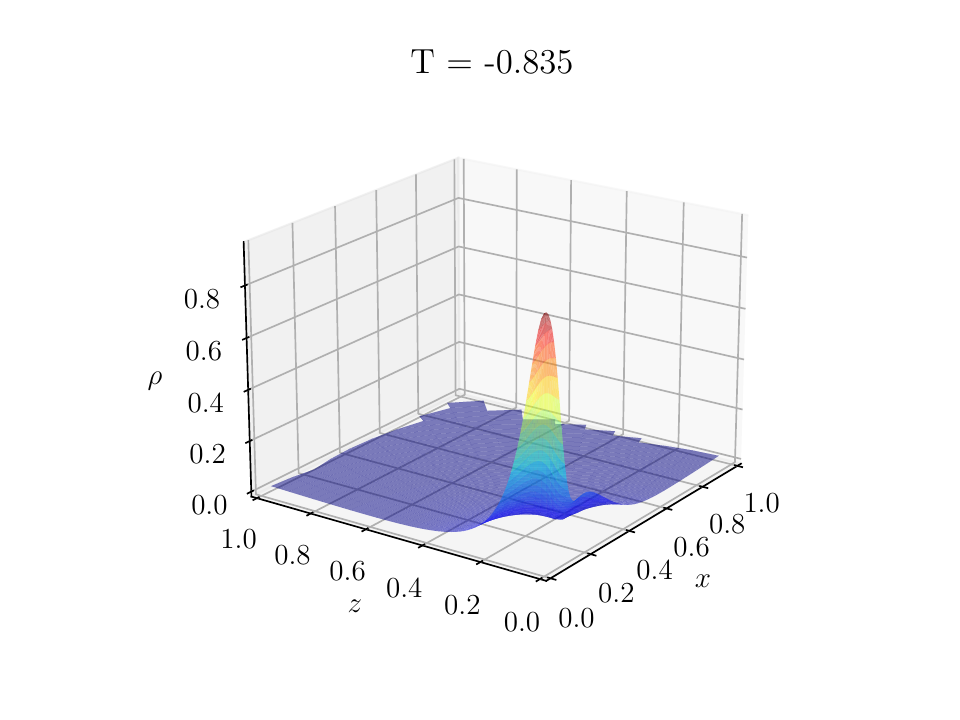}
     \includegraphics[width = 0.45 \textwidth]{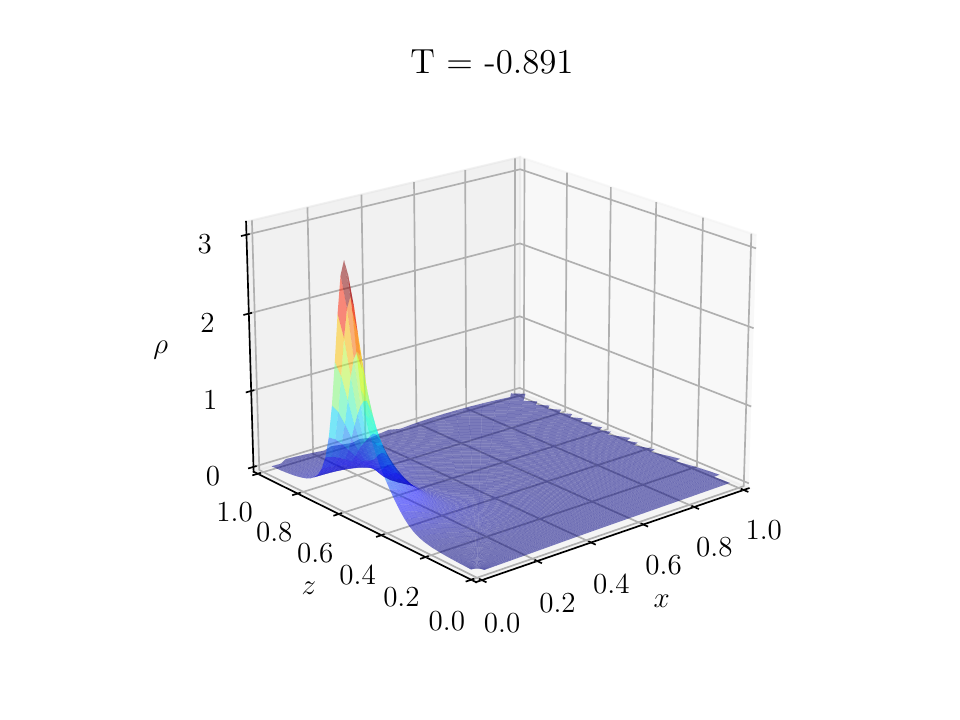}

     \includegraphics[width = 0.45 \textwidth]{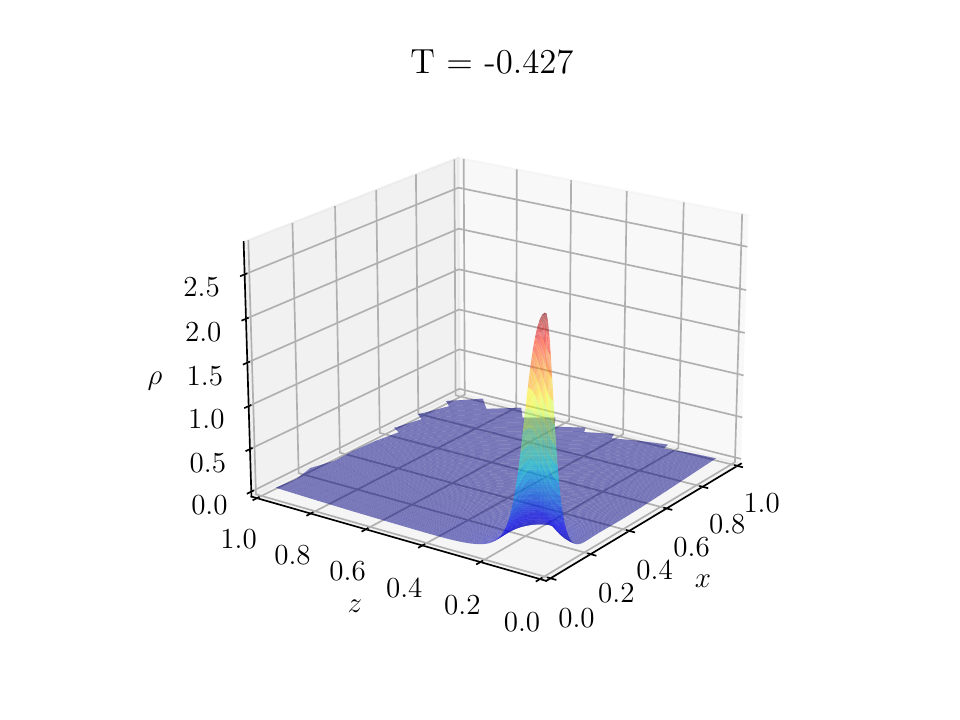}
     \includegraphics[width = 0.45 \textwidth]{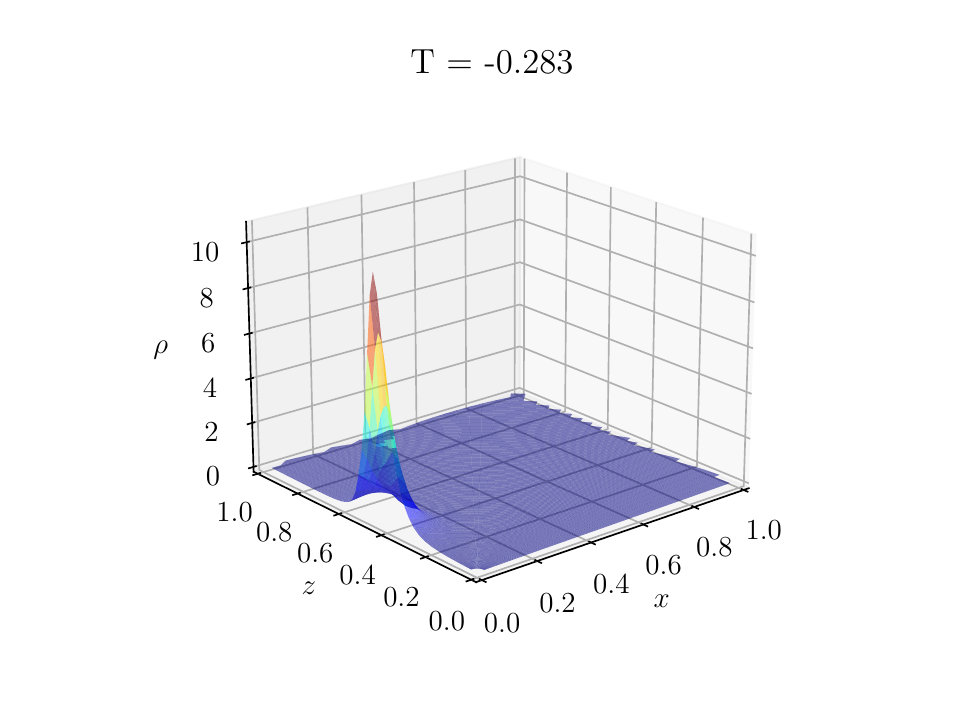}

     \includegraphics[width = 0.45 \textwidth]{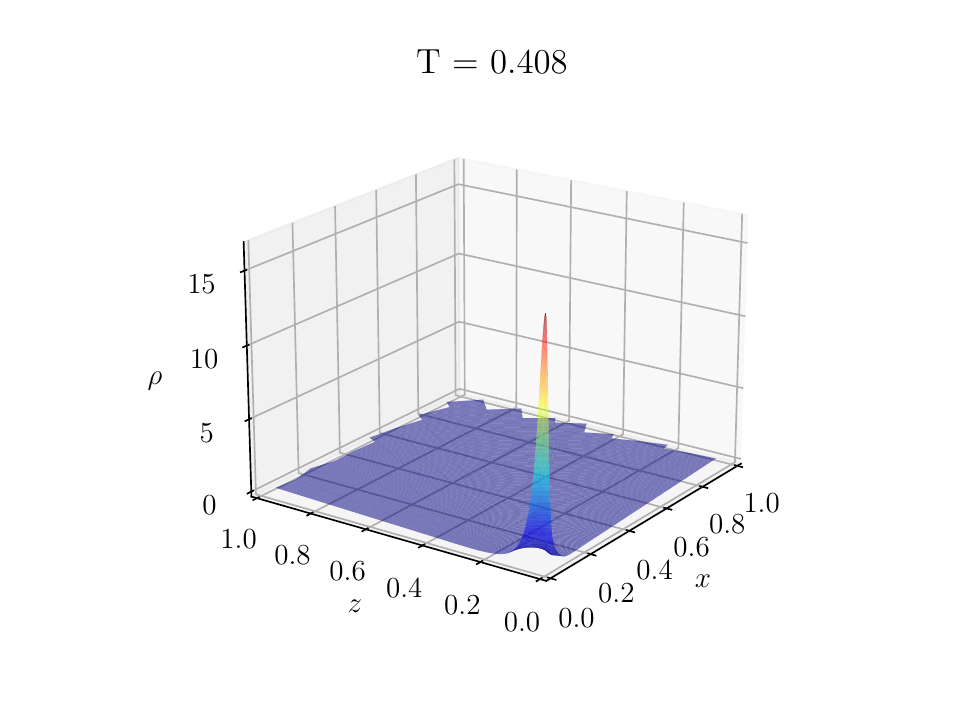}
     \includegraphics[width = 0.45 \textwidth]{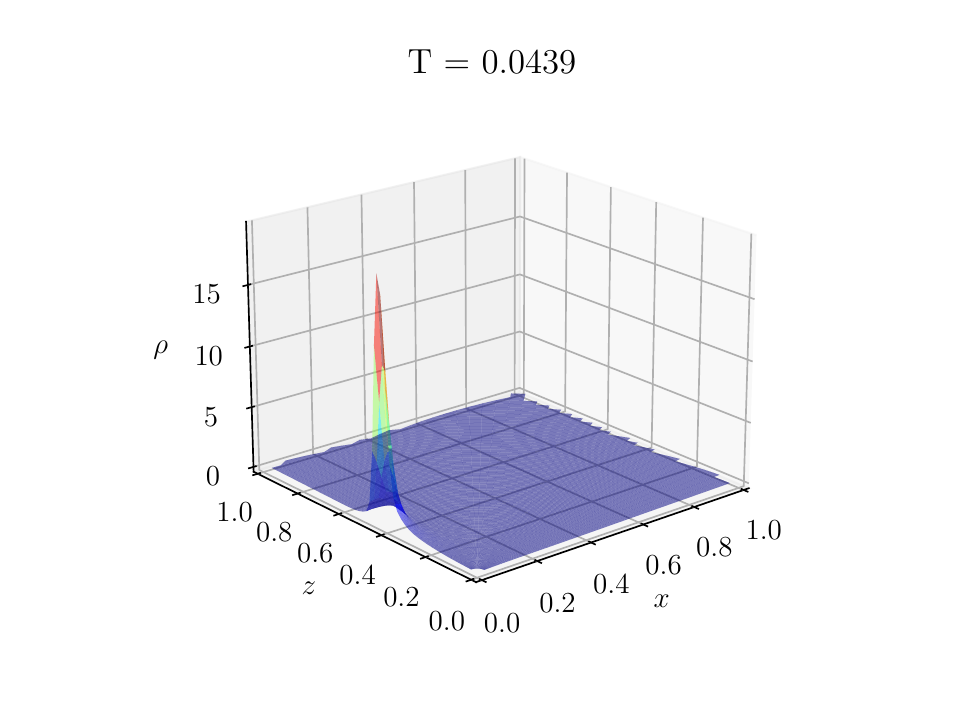}
    \caption{Same as Figs.~\ref{fig:Axi} and \ref{fig:lapse}, but for the density $\rho$.}
    \label{fig:rho}
\end{figure*}

\begin{figure}
    \centering
    \includegraphics[width = 0.5 \textwidth]{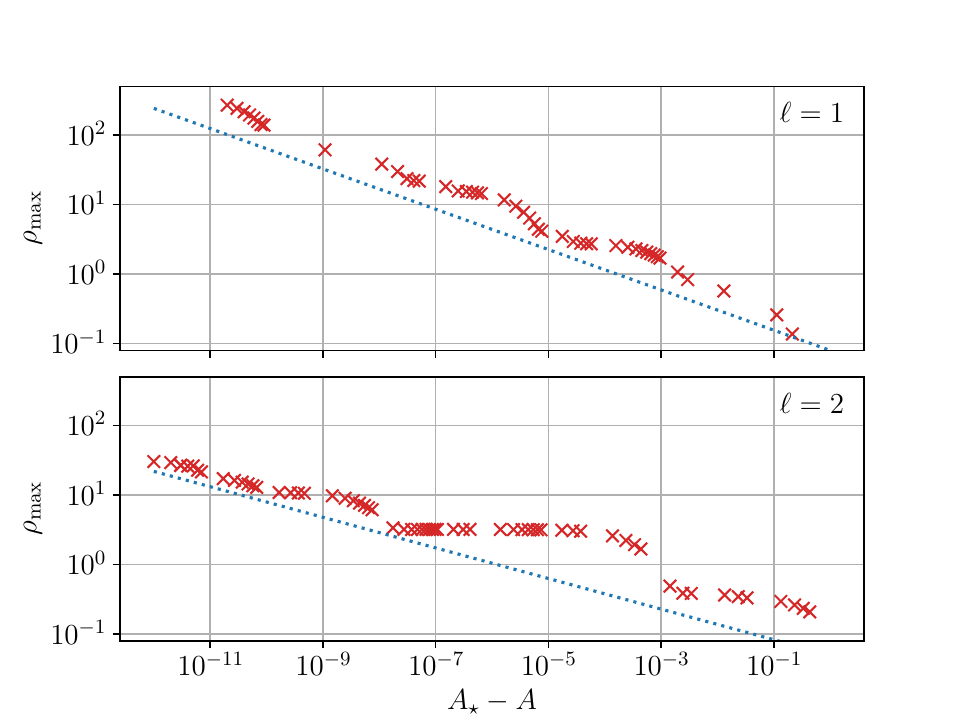}
    \caption{The maximum densities encountered for dipole (top panel) and quadrupole waves (bottom panel) as a function of $\mathcal{A}_\star - \mathcal{A}$.  The dotted lines are fits $\rho_{\rm max} \simeq (\mathcal{A}_\star - \mathcal{A})^{2 \gamma}$ with $\gamma^{\rm dip} = 0.145$ for the dipole waves (see BGH) and $\gamma^{\rm quad} = 0.11$ for the quadrupole waves. }
    \label{fig:max_dens_A}
\end{figure}

For subcritical data, the lapse function approaches unity at late times, as the wave disperses and leaves behind flat space.  For supercritical data, on the other hand, the lapse drops to zero at late times.  While the lapse is a coordinate-dependent quantity, other simulations of critical collapse with 1+log slicing have shown that such a ``collapse of the lapse" is indeed indicative of black-hole formation (see, e.g., \cite{HilBWDBMM13,BauM15,Bau18}, compare also with \cite{AkbC15}).

Note from Fig.~\ref{fig:lapse_of_t} that, for dipole waves, the dark and faded lines overlap for most of the evolution, indicating that the lapse takes its minimum value at the center (see also Fig.~\ref{fig:lapse} below).  This is consistent with the findings of BGH, who provided evidence for a critical solution with an accumulation point at the center for dipole waves.  For most of the quadrupole evolution, however, the lapse takes a minimum away from the center, including during the ``collapse of the lapse" for supercritical solutions.  This is a first suggestion that, for multipole moments higher than dipoles, centers of collapse form away from the center -- similar to the ``bifurcations" reported by \cite{ChoHLP03b,HilWB17,Bau18,LedK21}.  This finding may not be surprising, since we might expect centers of collapse at the locations of the highest densities.  For dipoles, these can be found at the center, but for higher multipole waves the energy density $\rho$ vanishes at the center (see Section \ref{sec:EMwaves}).

Fine-tuning the dipole data to about 11 digits results in quite short oscillation periods late in the evolution (see the top panel in Fig.~\ref{fig:lapse_of_t}), indicating that the evolution follows the critical solution until quite close to the accumulation event.  Fine-tuning the quadrupole data to the same number of digits, on the other hand, does not result in nearly as short oscillation periods (see the bottom panel in Fig.~\ref{fig:lapse_of_t}), meaning that the solution remains significantly further away from the accumulation event.  For the dipole data we can therefore estimate the proper time of the accumulation event rather accurately, $\tau_\star^{\rm dip} \simeq 5.66$, but for the quadrupole data this is much more difficult.  In the following we will adopt the value $\tau_\star^{\rm quad} \simeq 29.5$, which resulted in scaling behavior close to that expected for self-similar contraction (see Figs.~\ref{fig:rho_of_T}, \ref{fig:rho_prop_R} and \ref{fig:rho_photon}).  We caution, however, that this estimate is rather crude, and that the relative error in this value may be as large as 10\% or so; similar uncertainties affect all other values that we report in the following.

In Fig.~\ref{fig:rho_of_T} we show the maximum and central density $\rho$ (see Eq.~\ref{rho}) as a function of the ``slow time"
\begin{equation} \label{slow_time}
    T \equiv - \log(\tau_\star - \tau) + T_0,
\end{equation}
where $\tau$ is again the proper time of an observer at the origin, and where we have chosen the arbitrary offset $T_0$ to vanish for the dipole data, and $T_0 = 2$ for the quadrupole data.  We note that there is some ambiguity in how to best define $T$ when the centers of collapse are not at the origin; as an alternative to adopting the proper time of an observer at the origin, one could consider an observer whose worldline passes through those centers (see also the discussion in \cite{LedK21}).  The dotted lines in Fig.~\ref{fig:rho_of_T} represent curves proportional to $e^{2T} \propto (\tau_\star - \tau)^{-2}$, and hence the expected growth rate of the density in a self-similarly contracting solution.

As observed already by BGH, the evolution for dipole data is consistent with an approximate DSS critical solution; while this self-similarity is certainly not exact, the maxima in the density grow approximately at the expected rate $e^{2T}$, and it is possible to identify $\Delta^{\rm dip} \simeq 0.55$ as an approximate echoing period of the DSS critical solution.\footnote{Recall that the density $\rho$ is quadratic in the dynamical field $A_\xi$, and that the periodicity refers to that of the latter.}  While departures from an exact self-similarity are even larger for the quadrupole data, we again observe an over-all growth that is not inconsistent with the expected rate.  Moreover, we can again identify a dominant oscillation in the quadrupole data, and can estimate these oscillations to have a period of approximately $\Delta^{\rm quad} \simeq 0.3$.  It is difficult to determine this period of the DSS critical solution accurately, not only because of the departures from an exact periodicity, but also because of the ambiguities in the definition of the slow time (\ref{slow_time}) that we discussed above, and because we can determine $\tau_\star^{\rm quad}$ only crudely.  Despite these uncertainties, our findings suggest that the period $\Delta^{\rm quad}$ is shorter than $\Delta^{\rm dip}$, possibly by a factor of two.

\subsection{Profiles}

We next show profiles of some characteristic functions for near-critical evolutions at the times marked by the solid vertical (orange) lines in Figs.~\ref{fig:lapse_of_t} and \ref{fig:rho_of_T}, i.e.~at times at which the maximum density $\rho_{\rm max}$ on a spatial slice takes a (local) maximum in time.   In Figs.~\ref{fig:Axi}, \ref{fig:lapse}, and \ref{fig:rho} we compare profiles for dipole data in the left column with those for quadrupole data in the right column. 

We start in Fig.~\ref{fig:Axi} with profiles of the vector potential.  Specifically, we show profiles of the gauge-invariant quantity
\begin{equation} \label{Axi}
    A_\xi \equiv \frac{\xi^a A_a}{(\xi^a \xi_a)^{1/2}} = \frac{A_\varphi}{g_{\varphi\varphi}},
\end{equation}
which is formed from the vector potential $A_a$ and the Killing vector generating axisymmetry, $\xi^a$.  For the dipole data in the left column, the vector potential is symmetric across the equator (shown as the $x$-axis; see Section \ref{sec:dipole}) and takes a maximum there, while for the quadrupole data in the right column it is antisymmetric across the equator (see Section \ref{sec:quadrupole}).  For the quadrupole data, $A_\xi$ vanishes both on the equator and on the symmetry axis (shown as the $z$-axis); note that this results in large gradients close to the symmetry axis at late times.  

As a different way of presenting the same data we also show contour plots of $A_\xi$ in Fig.~\ref{fig:Axi_contour}.  In this plot, dashed lines represent contours of the dipole waves, while solid lines represent contours of quadrupole waves.  Each panel in the figure represents the data in the corresponding row of Fig.~\ref{fig:Axi}.  While it is not clear how exactly to identify a particular instant of the dipole evolution with one of the quadrupole evolution, the three chosen times appear to represent the respective evolutions at similar stages -- at least in terms of the spatial coordinates chosen in our simulations.

In Fig.~\ref{fig:lapse} we show profiles of the lapse function $\alpha$, again for near-critical evolutions.  As expected from Fig.~\ref{fig:lapse_of_t}, the lapse takes its minimum value at the center for the dipole data shown in the left column, while it takes a minimum value away from the center for the quadrupole data in the right column.  Note also that the minima become sharper at later times, which is consistent with a self-similar contraction.

Finally, we show profiles of the density $\rho$ in Fig.~\ref{fig:rho}.  Consistent with our earlier observations we notice that the density takes its maxima at the center for dipole data, and away from the center, on the symmetry axis, for quadrupole data.  As expected, the values of these maxima increase as time advances, and the density profiles become increasingly sharp.  The spherical polar coordinates of our code are ideally suited to resolve the density peaks when they occur at the center, i.e.~for the dipole data.  They are not well suited, however, to resolve density peaks away from the center, as for the quadrupole data.  Evidently the numerical resolution of those peaks becomes increasingly poor in our simulations.

\subsection{Scaling}

In Fig.~\ref{fig:max_dens_A} we graph the (global) maximum densities $\rho_{\rm max}$ encountered in simulations for given amplitudes $\mathcal{A}$ of the initial data 
(see Eqs.~\ref{DipoleEField} and \ref{QuadrupoleEField}), versus $\mathcal{A}_\star - \mathcal{A}$, where $\mathcal{A}_\star$ is the approximate critical value.    In Fig.~\ref{rho_scaling} we have adopted $\mathcal{A}_\star^{\rm dip} \simeq 0.91295765109$ and $\mathcal{A}_\star^{\rm quad} \simeq 3.533437407467$.  We also included, as the dotted lines, the expected power-law scaling 
\begin{equation} \label{rho_scaling2}
    \rho_{\rm max} \simeq (\mathcal{A}_\star - \mathcal{A})^{-2\gamma}
\end{equation}
(see Eq.~\ref{rho_scaling}), with fitted values of $\gamma^{\rm dip} = 0.145$ (see BGH) and $\gamma^{\rm quad} = 0.11$.  

For critical solutions that are DSS, one would expect a periodic ``wiggle" superimposed on the scaling (\ref{rho_scaling2}).  The absence of such a strictly DSS critical solution for electromagnetic waves is reflected by the absence of a strictly periodic wiggle in Fig.~\ref{fig:max_dens_A}.  We nevertheless observe a general trend in the data that is not inconsistent with a power-law scaling of the form (\ref{rho_scaling2}).  While the results for dipole data are based on simulations that resolve the solution well even close to the black hole threshold, we expect our numerical results for quadrupole data to be affected by the lack of sufficient resolution away from the center, and therefore to be less reliable.  Our results nevertheless suggest that the critical exponent for the quadrupole data is different from that for dipole data.  This is consistent with the finding of \cite{LedK21}, who reported critical exponents for vacuum gravitational wave collapse that also depend on the choice of initial data.

\begin{figure}
    \centering
    \includegraphics[width = 0.5 \textwidth]{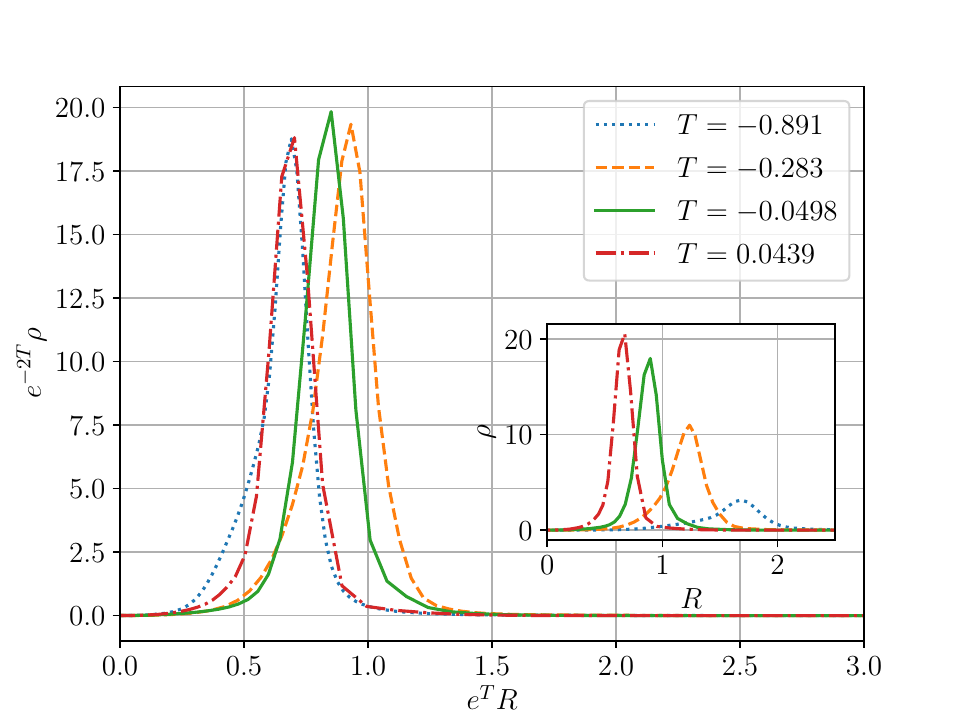}
    \caption{The density $\rho$ along the axis of symmetry (i.e.~the $z$-axis in Fig.~\ref{fig:rho}) as a function of proper distance $R$ from the center, for a near-critical quadrupole evolution.  We show results for the four different times marked by all four vertical lines in Figs.~\ref{fig:lapse_of_t} and \ref{fig:rho_of_T}, including the three times shown in the snapshots of Fig.~\ref{fig:rho}.  The inset shows the raw data, while the main plot shows the data rescaled according to the expectation for a self-similar contraction.}
    \label{fig:rho_prop_R}
\end{figure}

\begin{figure}
    \centering
    \includegraphics[width = 0.5 \textwidth]{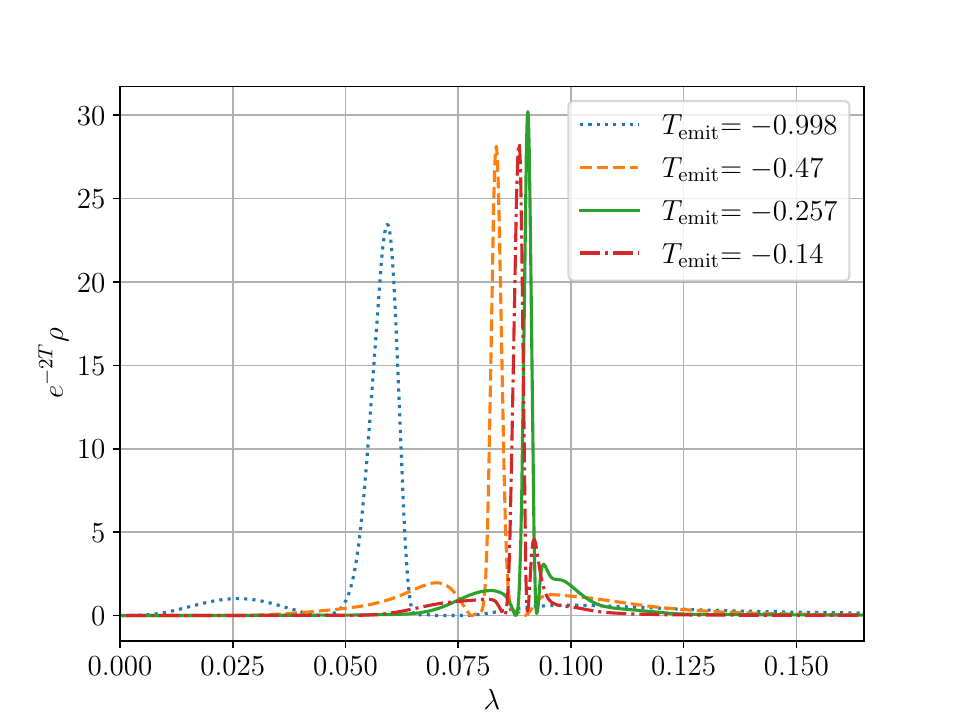}
    \caption{The density $\rho$ along the trajectories of null-geodesics emitted from the center at slow times $T_{\rm emit}$, chosen so that the geodesics pass through the same density peaks as those shown in  Fig.~\ref{fig:rho_prop_R}.}
    \label{fig:rho_photon}
\end{figure}

\subsection{Uniqueness of the critical solution}

It is clear from both our discussion in Section \ref{sec:EMwaves} as well as Figs.~\ref{fig:Axi} -- \ref{fig:rho} that the critical solution cannot be unique {\em globally}.  At least in principle, however, it is possible that the critical solution remains unique {\em locally}.  For quadrupole data, this might be the case if each one of the centers of collapse behaved just like that for dipole data: while the critical solution would differ globally, it might be very similar in the vicinity of each center of collapse.  Referring to the appearance of two centers of collapse as a ``bifurcation" might suggest exactly that -- namely that the new off-center centers of collapse are indeed such ``copies" of dipole center of collapse, with very similar properties.  

There is some evidence, however, that suggests otherwise.  We first observe from Fig.~\ref{fig:rho_of_T} that the echoing period $\Delta^{\rm quad}$ for quadrupole waves appears different from that for dipole waves, $\Delta^{\rm dip}$. As we discussed in Section \ref{sec:lapse_dens}, there is some ambiguity even conceptually in how to define $T$ for quadruole waves, and hence $\Delta^{\rm quad}$.  Our results nevertheless suggest that $\Delta^{\rm quad}$ is smaller than $\Delta^{\rm dip}$, as one might expect if the oscillations are indeed associated with a higher-order mode.  If, on the other hand, the centers of collapse for quadrupole waves had the same properties as that for dipole waves, we would expect to observe the same period $\Delta$ for both.

Similarly, if the centers of collapse for quadrupole waves were ``copies" of the single center of collapse for dipole waves, one would expect the critical exponent $\gamma$ to be identical.  Our data, shown in Fig.~\ref{fig:max_dens_A}, instead suggest that $\gamma^{\rm quad}$ is somewhat smaller than $\gamma^{\rm dip}$.  While this result may well be be affected by numerical error resulting from the poor numerical resolution of the off-center peaks on our spherical grids, it is consistent with the findings of \cite{LedK21}, who similarly found different critical exponents for different families of initial data in simulations of critical collapse of vacuum gravitational waves.

If, on the other hand, the two centers of collapse are indeed features of a distinct critical solution, then one would expect the distance between the two peaks to scale together with the rest of the solution.  While the two peaks do appear to approach each other in Fig.~\ref{fig:Axi_contour}, for example, we caution that the spatial coordinates shown there have no immediate physical meaning.  We therefore show in Fig.~\ref{fig:rho_prop_R} the density $\rho$ along the symmetry axis as a function of proper distance $R$ from the center at the four times marked by the vertical lines in Figs.~\ref{fig:lapse_of_t} and \ref{fig:rho_of_T}.  The inset shows the ``raw" data, while the large panel shows both density and proper distance rescaled assuming a self-similar contraction.  Clearly the agreement of the rescaled quantities at different times is not perfect -- and in the absence of a strict self-similarity we cannot expect that -- but evidently the agreement of the rescaled quantities is significantly better than that of the raw data.

In Fig.~\ref{fig:rho_prop_R}, proper distance was measured along a spatial slice, which makes this distance a slicing-dependent quantity.  Alternatively, we consider null geodesics that propagate along the symmetry axis, emitted from the center of symmetry at (slow) times $T_{\rm emit}$.  For each null geodesic we introduce an affine parameter $\lambda$ normalized such that $\lambda = 0$ at the center, and $d\lambda/dt$ measured along the null-geodesic equal to $dT/dt$ at the center.  With this normalization, the parameter $\lambda$ ``inherits" the natural scaling of the self-similar solution.  

In Fig.~\ref{fig:rho_photon} we show graphs of rescaled densities $e^{-2T} \rho$ as a function of $\lambda$ for such null-geodesics, chosen such that they pass through the same peaks in the density as those shown in Fig.~\ref{fig:rho_prop_R}.  While the last three peaks again agree reasonably well, the first peak shows a larger deviation, but that might be caused by its null geodesic having been emitted from the center before the solution enters its approximately self-similar stage.  We again conclude that our results are not inconsistent with a self-similar decrease in the distance between the two peaks -- suggesting that the two centers of collapse might be features of a global critical solution for quadrupole waves, rather then ``copies" of the center of collapse encountered for dipole waves.

\section{Summary and Discussion}
\label{sec:summary}

In this paper we study critical phenomena in the gravitational collapse of electromagnetic waves.  Generalizing results of BGH, who focused on dipole initial data, we also consider quadrupole initial data and find several qualitative differences.  Most importantly, we observe that dipole data feature a single center of collapse at the center of symmetry, but quadrupole data feature a pair of centers of collapse on the symmetry axis, above and below the center of symmetry.  A similar ``bifurcation" has previously been reported both for (non-spherical) scalar fields \cite{ChoHLP03b,Bau18} and gravitational waves \cite{HilWB17,LedK21}.  This observation alone demonstrates that the critical solution for electromagnetic waves cannot be unique, at least not {\em globally}.  Actually, this conclusion follows already from the fact that electromagnetic waves with odd $\ell$ are symmetric across the equator, while those with even $\ell$ are antisymmetric (see Section \ref{sec:EMwaves}).  The absence of a unique critical solution may be a general feature of critical collapse in cases that do not allow a spherically symmetric critical solution; this, in fact, has also been suggested by the toy model presented in \cite{SuaVH21}.

The above conclusion leaves open the possibility that the critical solution might be unique {\em locally}, in the sense that the two centers of collapse observed for quadrupole data, for example, might be ``copies" of that found for dipole data.  In fact, referring to the appearance of two centers of collapse as a ``bifurcation" might suggest such a behavior.  We provide some evidence to the contrary, however, namely that the critical solution for quadrupole data is distinct from that for dipole data even locally.  In particular, we observe different echoing periods and critical exponents for the different multipole moments; some of these observations appear to be consistent with results for gravitational wave collapse (e.g.~\cite{HilWB17,LedK21}).  We also find that the distance between the centers of collapse found for quadrupole waves appears to scale in a manner that is not inconsistent with an approximately self-similar contraction.  Together, these findings suggest that the two centers of collapse found for quadrupole waves might be features of a global critical solution for quadrupole waves, rather than two distinct local copies of the dipole critical solution.

Given the similarities between some of our observations and those for critical collapse of gravitational waves (e.g.~\cite{HilWB17,LedK21}) we speculate that, for both electromagnetic and gravitational waves, the absence of a spherically symmetric critical solution might lead to the absence of a unique critical solution (see also \cite{SuaVH21}).  Fine-tuning a given family of initial data may, in both cases, lead to an approximately self-similar critical solution, with associated approximate scaling and critical exponents, but they may be different for different families of initial data.

As we discussed in Section \ref{sec:indata}, multipoles of odd (even) order $\ell$ will couple gravitationally to other modes of odd (even) order.  This coupling may lead to a ``competition" between different modes of odd (even) order, not unlike the competition between a scalar field and a Yang-Mills field as discussed in \cite{GunBH19}.  While scalar and Yang-Mills fields have two distinct critical solutions individually, the authors of \cite{GunBH19} found that, for sufficient fine-tuning, the scalar field always dominates, so that, on sufficiently small scales, the critical solution becomes unique again.  It is possible that the coupling between all electromagnetic modes of either odd or even order leads to a similar competition, and it is further possible that, with sufficient fine-tuning, one such mode will again dominate on sufficiently small scales (ignoring any additional competition between electromagnetic and gravitational degrees of freedom).  If so, this would result in the emergence of only two critical solutions: one for which the electromagnetic fields are symmetric across the equator, and a second one for which they are antisymmetric.

\acknowledgements

It is a great pleasure to thank Carsten Gundlach and David Hilditch for numerous elucidating discussions, as well as a careful reading of a draft of this paper.  MFPM acknowledges support through an undergraduate research fellowship at Bowdoin College, and would like to thank Chloe Richards for many helpful conversations.  Numerical simulations were performed on the Bowdoin Computational Grid.  This work was supported in parts by National Science Foundation (NSF) grants PHY-1707526 and PHY-2010394 to Bowdoin College.

\begin{appendix}
\section{Analytical solutions to Maxwell's equations in flat vacuum spacetimes}
\label{sec:ana_sols}

In order to derive the analytical solutions of Section \ref{sec:EMwaves} we first combine Maxwell's equations (\ref{maxwell}) into a single equation for the vector potential $A^a$,
\begin{equation} \label{app_max1}
    - \partial_t^2 A^a + \nabla^b \nabla_b A^a - \nabla^a \nabla_b A^b = 0,
\end{equation}
where, as before, we have assumed vacuum.  Further assuming axisymmetry in a flat spacetime, and adopting spherical polar (Minkowski) coordinates, we focus on solutions for which only the $A^{\hat \varphi}$ component is non-zero; Eq.~(\ref{app_max1}) can then be written as
\begin{equation} \label{app_max2}
    - \partial_t^2 \tilde A + \partial_r^2 \tilde A + 
    \frac{1}{r^2 \sin \theta} \partial_\theta \left( \sin \theta \, \partial_\theta \tilde A \right) - \frac{\tilde A}{r^2 \sin^2 \theta} = 0
\end{equation}
where we have defined $\tilde A = r A^{\hat \varphi}$.  We now look for separable solutions of the form
\begin{equation} \label{app_sep_ansatz}
\tilde A(t, r, \theta) = g(t,r) \, f(\theta).
\end{equation}
Inserting this ansatz into (\ref{app_max2}) shows that the 
angular functions $f(\theta)$ have to satisfy
\begin{equation} \label{app_angular}
    \frac{1}{\sin \theta} \partial_\theta\left( \sin \theta \, \partial_\theta f \right) - \frac{f}{\sin^2 \theta} = - \ell ( \ell + 1) f,
\end{equation}
where $\ell$ is a constant, while the time-radial functions $g(t,r)$ satisfy
\begin{equation} \label{app_radial}
    - \partial_t^2 g + \partial_r^2 g - \frac{\ell(\ell + 1)}{r^2} \, g = 0,
\end{equation}
a special case of the Euler-Poisson-Darboux equation.

Regular solutions $f_\ell(\theta)$ to (\ref{app_angular}) exist if $\ell$ is a positive integer; these solutions are related to the components of the axisymmetric magnetic vector spherical harmonics.  In the following we adopt
\begin{equation} \label{app_fs}
\begin{array} {rcll}
    f_1(\theta) & = & \sin \theta & (\ell = 1) \\
    f_2(\theta) & = & \cos \theta \sin \theta & (\ell = 2) \\
    f_3(\theta) & = & ( 5 \cos^2 \theta - 1) \, \sin \theta~~~~~~ & (\ell = 3) 
\end{array}
\end{equation}
for dipole, quadrupole, and octupole waves.   Solutions to (\ref{app_radial}) can be then constructed with the ansatz
\begin{equation} \label{app_radial_ansatz}
g_\ell(t,r) = \sum_{j = 0}^\ell c_j r^{j-l} \, F_{\pm}^{(j)}(x),
\end{equation}
where the $c_j$ are constants and where $F_\pm^{(j)}(x) = d^j F_\pm(x)/dx^j$ is the $j$-th derivative of a function $F_\pm(x)$ of $x = r \pm t$ describing ingoing (``+") or outgoing (``-") waves (see, e.g., \cite{Rin08}).

We can construct dipole waves by adopting $\ell = 1$ in the above expressions, in which case the ansatz (\ref{app_radial_ansatz}) reduces to
\begin{equation} \label{app_g1}
    g_1 = \frac{c_0 F_\pm}{r} + c_1 F_\pm^{(1)}.
\end{equation}
Inserting this into (\ref{app_radial}) yields $c_1 = - c_0$.  We choose $c_0 = 1$, so that (\ref{app_g1}) reduces to
\begin{equation}
    g_1 = \frac{F_\pm}{r} - F_\pm^{(1)}.
\end{equation}
We may then assemble the dipole solution for $A^{\hat \varphi}$ from
\begin{equation}
A^{\hat \varphi}_\pm = \frac{g_1 f_1}{r} = \left( \frac{F_\pm}{r^2} - \frac{F_\pm^{(1)}}{r} \right) \sin \theta.~~~~~~~~~(\ell = 1)
\end{equation}
A time-symmetric solution can be constructed from a superposition of ingoing and outgoing waves,
\begin{equation}
    A^{\hat \varphi} = A^{\hat \varphi}_- - A^{\hat \varphi}_+
\end{equation}
with $F_+ = F_-$.  Choosing regular functions $F_\pm$ that are even in $x$ will then result in regular solutions for the vector potential $A^{\hat \varphi}$.  In particular, the Gaussian profile
\begin{equation}
    F_\pm = {\mathcal A}\, \sigma^2 e^{-(r \pm t)^2/\sigma^2}
\end{equation}
yields the dipole solution (\ref{DipoleSolution}).

Quadrupole ($\ell = 2$) and octupole ($\ell = 3$) waves can be constructed similarly.  Specifically, we find 
\begin{equation}
    g_2 = \frac{F_\pm}{r^2} - \frac{F_\pm^{(1)}}{r} + \frac{F_\pm^{(2)}}{3}
\end{equation}
for quadrupole waves and 
\begin{equation}
    g_3 = \frac{F_\pm}{r^3} - \frac{F_\pm^{(1)}}{r^2} + \frac{2 F_\pm^{(2)}}{5 r} - \frac{F_\pm^{(3)}}{15}
\end{equation}
for octupole waves.  Combining these with the respective angular functions $f_2$ and $f_3$ in (\ref{app_fs}), using a superposition of ingoing and outgoing waves, and choosing Gaussian profiles for $F_\pm$ then yields the quadrupole waves (\ref{QuadrupoleSolution}) and the octupole waves (\ref{OctupoleSolution}).

\end{appendix}

%


\end{document}